\documentclass[aps,prb,twocolumn,superscriptaddress,floatfix]{revtex4-1}
\usepackage{amsmath,amssymb,graphicx,graphics,color}
\usepackage{dcolumn}
\usepackage{bm}
\usepackage[normalem]{ulem}
\usepackage{psfrag}
\newcommand{\ket}[1]{|#1\rangle}
\newcommand{\bra}[1]{\langle#1|}

\newcommand{\bs}{\boldsymbol}
\newcommand{\ignore}[1]{}

\def\im{{\sf {i}}}

\def\one{{\mathbb{1}}}

\def\one{{\mathchoice {\rm 1\mskip-4mu l} {\rm 1\mskip-4mu l} {\rm 1\mskip-4.5mu l} {\rm 1\mskip-5mu l}}}

\begin{document}

\title{Phase transitions in the $\mathbb{Z}_p$ and U(1) clock models}

\author{G. Sun }
\affiliation{ College of Science, Nanjing University of Aeronautics and Astronautics, Nanjing, 211106, China}

\author {T. Vekua}
\affiliation{Department of Physics, Indiana University, Bloomington, IN 47405, USA}

\author{E. Cobanera}
\affiliation{Department of Mathematics and Physics, SUNY Polytechnic Institute, 100 Seymour Rd, Utica, NY 13502, USA }

\author {G. Ortiz}
\affiliation{Department of Physics, Indiana University, Bloomington, IN 47405, USA}

\begin{abstract}
Quantum phase transitions are studied in the non-chiral $p$-clock chain, and a new explicitly
U(1)-symmetric clock model, by monitoring the ground-state fidelity susceptibility. For 
$p\ge 5$, the self-dual $\mathbb{Z}_p$-symmetric chain displays a double-hump structure in the fidelity susceptibility
with both peak positions and heights scaling logarithmically to their corresponding 
thermodynamic values. This scaling is precisely as expected for two Beresinskii-Kosterlitz-Thouless 
(BKT) transitions located symmetrically about the self-dual point, and so confirms numerically the
theoretical scenario that sets \(p=5\) as the lowest \(p\) supporting BKT transitions in 
$\mathbb{Z}_p$-symmetric clock models. For our U(1)-symmetric, non-self-dual minimal modification
of the $p$-clock model we find that the phase diagram depends strongly on the parity 
of $p$ and only one BKT transition survives for $p\geq 5$. 
Using asymptotic calculus we map the self-dual clock model exactly, in the large 
$p$ limit, to the quantum $O(2)$ rotor chain. Finally, using bond-algebraic dualities we estimate the critical 
BKT transition temperatures of the classical planar $p$-clock models defined on square lattices, 
in the limit of extreme spatial anisotropy. Our values agree remarkably well with those determined
via classical Monte Carlo for isotropic lattices. This work highlights the power of 
the fidelity susceptibility as a  tool for diagnosing the BKT transitions 
even when only discrete symmetries are present. 

\end{abstract}

\maketitle

\date{\today}


\section{Introduction}

Global continuous symmetries cannot undergo spontaneous symmetry breakdown in classical 
two-dimensional systems at finite temperature \cite{M-W,reviewMW,book}. In one-dimensional quantum 
systems, they cannot be broken even at zero temperature (ground state) in an effectively Lorentz 
invariant way \cite{Coleman73}. As a result, many low-dimensional models of magnetism do not actually 
display magnetic ordering: the magnetization, their natural order parameter, vanishes 
in the thermodynamic limit for any amount of fluctuations however small. 
This fact does not preclude the existence of phase transitions, which are associated to non-analyticities of the 
thermodynamic state. Such transitions, however,  must 
necessarily be beyond the scope of Landau's theory
which relies on the existence of a local order parameter \cite{book}.
The classical two-dimensional XY model is the textbook example of a system with no
magnetic ordering and yet hosts two distinct thermodynamic phases \cite{book}, one gapless (critical) and
the other gapped. 
$\!$Berezinskii, Kosterlitz and Thouless (BKT) \cite{Berezinskii,KT, K, book,book40years}
realized that the  transition separating these two phases is topological in nature: 
Bound vortex-antivortex pairs in the low-temperature critical phase become deconfined 
in the high-temperature phase. Those topologically-charged vortices are responsible for 
the exponential decay of spin-spin correlations  at high temperature. 

As the BKT transition of the XY model became more widely known,
several other classical and quantum phase transitions (QPTs) were identified as BKT. In fact, 
a survey of experimental data [\onlinecite{Taroni2008}] 
suggests that the BKT transition is ubiquitous in two dimensions, closely followed by the
Ising class, while the classes of the three and four-state Potts models are far less common. 
At the most basic level, extrapolating naively from the XY model, 
the BKT transition is somehow the result of having a global U(1) symmetry in 
classical (quantum) planar (chain) systems, so that the Mermin-Wagner-Coleman theorems 
\cite{book}  will preclude the emergence of an order parameter, and yet there is a local field
(e.g., the local magnetization) that winds around loops to yield topologically-charged
vortices. Consistent with this picture, a  prototype of a BKT transition in quantum
systems is the  superfluid to Mott-insulating phase transition of the  
Bose-Hubbard chain at integer fillings\cite{Sachdev}.
But as one tries carefully to apply the qualifications of BKT to phase transitions in models beyond the 
planar classical XY model or its quantum incarnation, the quantum chain of planar rotors,
conceptual complications quickly arise. Consider, for instance, the issue of the  
nonexistence of a local order parameter. This is true of the XY model. However,
there are models where a local
order parameter distinguishing the two phases exists and yet the phases are 
separated by a BKT transition. 
This is true for example of the QPT separating the U(1)-symmetric
critical phase from the antiferromagnetic  
gapped phase of the half-integer-spin XXZ chain. The transition happens at the SU(2)-symmetric
point, and the staggered magnetization ($\mathbb{Z}_2$ order parameter) changes from a non-zero value to
vanishing at this BKT point and throughout the gapless phase. 

\begin{figure}
\includegraphics[width=8.0cm]{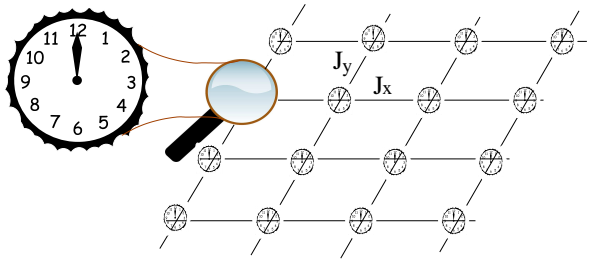}
\caption{The classical, planar (2$d$) $\mathbb{Z}_p$ clock model with $p=12$. }
\label{fig:clock}
\end{figure}

A different twist comes from systems with solely {\it discrete} 
global symmetries that are also believed to display BKT transitions. Two important examples are the \(p\)-clock
models, for sufficiently large $p$, \cite{Kadanoff, Elitzur, Fro,Ortiz} and the classical planar 
(quantum chain) anisotropic next-nearest-neighbor
(ANNNI) model \cite{ANNNI,ANNNIbook}. Are these models suggesting that a global U(1) symmetry
is not a necessary condition of a BKT transition? Not quite,
because such a symmetry could emerge at low energies \cite{Batista-Ortiz2004}.  
The $\mathbb{Z}_p$ clock models  (see Fig. \ref{fig:clock}) realize such a mechanism 
for large enough positive integer $p$\cite{Fro,Ortiz}. In fact,
for  the quantum \(p\)-clock it was shown in Ref.\,[\onlinecite{Ortiz}] 
that a continuous U(1) symmetry {\it emerges} for \(p\) as low as $p=5$ and this symmetry
is directly responsible for the existence of a critical phase for \(p\geq 5\). 
This critical phase, characterized by continuously varying critical exponents, is 
separated from the other two gapped phases of clock models with \(p\geq 5\) (a magnetically-ordered and a
paramagnetic phase) by BKT phase transitions. For $p\to \infty$, the classical planar $\mathbb{Z}_p$ clock 
model converges to the XY model, and the magnetically-ordered phase reduces to a point at 
temperature $T=0$. For a discussion of the critical phase and BKT transition in the ANNNI model, including
an indication of an emergent U(1) symmetry, see Ref.\,[\onlinecite{Allen2001}]. 

In summary, there are plenty of examples of BKT transitions, meaning loosely that
the transition of interest has some features in common with the phase transition of the XY model. 
We saw nevertheless that, in contrast to what one expects from the XY model, there may exists a local
order parameter that distinguishes the gapped from the gapless phase separated by a BKT point.
Similarly, while it seems generically possible to identify an exact or emergent global U(1)
symmetry in models with BKT transitions, there are plenty of models with U(1) symmetry,
gapless and gapped phases, and no BKT transition anywhere in their phase diagrams! We will discuss 
a few examples of such models shortly. 

 {What exactly makes a phase transition to be BKT?} Taking the XY model
as a defining paradigm, one can enumerate the not necessarily independent conditions that a 
critical point should satisfy to qualify as BKT: 

\begin{enumerate}

\item{A BKT criticality  
should always mark the boundary between a gapless {\it phase} and a gapped {\it phase}.} 

\item {The system should display an explicit or emergent (in the low-energy sector) global U(1) symmetry.}

\item{The free energy (classical) or ground state energy density (quantum) should 
display an essential singularity
at the BKT temperature, $T_{\sf BKT}$, or control parameter $\lambda_{\sf BKT}$ (infinite-order phase transition). }

\item {The fixed point action should be that of a Gaussian conformal field theory (GCFT) 
with universal critical exponents.  This feature, together with condition 1 above, is responsible 
for the universal jump of the helicity modulus across a BKT transition.}

\item {Non-universal critical exponents should change continuously in the critical phase as 
the BKT transition is approached. The deviation of the non-universal critical exponents from 
their universal values at the BKT point should develop a square-root behavior ($\sim \sqrt{|T_{\sf BKT}-T|}$ or 
$\sqrt{|\lambda_{\sf BKT}-\lambda|}$) when approaching the BKT point. }

\item{ There should appear multiplicative logarithmic corrections in correlation 
functions due to marginally irrelevant perturbations, and  not involving conserved currents of the underlying GCFT.}
\label{item6}

\item {There should be present a square-root singularity in the exponential divergence of the correlation length 
$\xi$ when approaching the BKT transition from the gapped side, i.e., $\log(\xi/a) \sim \! 1/ \sqrt{|T_{\sf BKT}-T|} $ or 
$1/ \sqrt{|\lambda_{\sf BKT}-\lambda|}$ 
, where $a$ is the lattice constant. }
\end{enumerate}
Table\,\ref{MainTable} summarizes a variety of prototypical models that satisfy some or all of these conditions. 

\begin{table*}
\centering
\begin{tabular}{|l|c|c|c|c|c|} \hline
 {\bf Quantum Chain Model} & {\bf Transition(s)} & {\bf Type}& {\bf Local OP} & {\bf Conditions} & 
 {\bf See} \\ \hline \hline
 {\sf XY (planar rotor)}           & confinement-deconfinement & {\bf BKT} & No & 1 - 7 & [\onlinecite{KT,book}]\\ \hline
 {\sf Bose-Hubbard } & superfluid-Mott insulator     & {\bf BKT} & No & 1 - 7  & [\onlinecite{Sachdev}] \\ \hline
  {\sf s=1/2 XXZ }    & critical-AF  & {\bf BKT} & Yes     &  1 - 7& [\onlinecite{KM}] \\ \hline
  {\sf ANNNI transverse field} &  critical-F & {\bf BKT} & Yes & 1 - 7 & [\onlinecite{Allen2001}] \\ \hline 
 {\sf $p$-clock, $p \ge 5$ }& critical-F, critical-Para &  {\bf BKT} &Yes, No & 1 - 7 & [\onlinecite{Fro,Ortiz}] \\ \hline \hline
{\sf s=1/2 XXZ }    & critical-F      & First Order & Yes & 1 , 2 & [\onlinecite{KM}]   \\   \hline                                    
 {\sf s=1/2 XX transverse field}  & critical - magnetized (gapped) & C-IC& Yes & 1 , 2 & [\onlinecite{KZDeng}] \\ \hline 
 {\sf s=1/2 $J_1$-$J_2$ SU(2)-symmetric AF} & critical - dimerized & Gaussian & Yes & 
1 - 4 & [\onlinecite{HaldaneJ1J2}]\\ \hline
{\sf s=1 Luther-Scalapino } &  critical-gapped & Gaussian & No & 1 - 4 & [\onlinecite{Luther-Scalapino}] \\ \hline
\end{tabular}
\caption{
Summary of representative models mentioned in this work. 
}
\label{MainTable}
\end{table*}

In this paper we analyze critically these conditions by way of density-matrix renormalization group (DMRG)
 simulations of quantum models chosen or designed to highlight the potential characteristics that 
may lead to a BKT transition. We start with the standard quantum 
self-dual $\mathbb{Z}_p$ clock model and then  introduce a new explicitly U(1)-symmetric clock model.
Our present study confirms the standard picture that for $p\ge 5$ quantum $p$-clock chains support two 
BKT phase transitions located symmetrically around the self-dual point \cite{Ortiz}. The phase diagram for the U(1) 
clock models is more exotic. By enhancing the $\mathbb{Z}_p$ to a U(1) symmetry we lose the 
self-duality symmetry of the phase diagram, but we gain in return a richer phase diagram,  
including critical phases and transitions that are not always BKT (see Fig.\,\ref{fig:u1sketch}).
We also map our critical coupling constants $\lambda_{\sf BKT}$ for quantum clock models to critical temperatures 
$T_{\sf BKT}$ of their two-dimensional classical counterparts using enhanced formulas 
for the quantum-to-classical mapping from
Ref.\,[\onlinecite{Ortiz}]. Our critical temperatures agree surprisingly well with those
calculated directly for the classical models with Monte Carlo techniques.   
Finally, we analyze by analytic means another paradigmatic quantum example of BKT phenomenology at 
any temperature including zero, the quantum rotor chain \cite{Sachdev}. 
By establishing a mathematically 
well-controlled large-$p$ limit of the self-dual $p$-clock chains, we show that the $O(2)$ quantum rotor model 
becomes indistinguishable from the \(p\)-clock model at low energies and sufficiently large \(p\). 
 
Our investigations are prompted by two important problems in the theory of critical phenomena in
two dimensions. First, there is certain degree of confusion in the literature on what should be 
called BKT topological transition (see Table \ref{MainTable}). There are well-established examples of phase transitions 
between gapped and gapless phases that should clearly not be called BKT transitions. 
For example, the ferromagnetic (F) transition in the XXZ half-integer spin chain, when 
changing the anisotropy parameter, satisfies conditions 1 and 2,  but it is of first order in nature. 
There are also examples of {\it continuous} phase transitions that satisfy 1 and 2 and
are not BKT transitions either. This is the case for the Lifshitz or commensurate-incommensurate (C-IC) transition 
of the XY spin-1/2 chain (with $J_x=J_y$) in transverse magnetic field at a saturation value \cite{KZDeng}. 
This kind of continuous transition does not meet some or all of the requirements 3 to 7, and it is not 
unusual in the literature to find authors that explicitly ignore conditions 5 to 7 in order to categorize
certain transitions as BKT. For example, the dimerization transition in 
the $J_1$-$J_2$ antiferromagnetic (AF) SU(2)-symmetric Heisenberg spin-1/2 chain, 
even though it does display an essential singularity and is sometimes characterized as a BKT transition, 
it does not satisfy conditions 5 to 7.  Rather, this critical point is described by an SU$_1(2)$ 
Wess-Zumino CFT (which is equivalent to Gaussian criticality) without the marginal corrections
that are responsible 
for the anomalous dimensions observed for non-conserved operators. Hence, there are no 
multiplicative logarithmic corrections in the correlation functions for the SU(2)-symmetric chain
at the dimerization transition point. However, logarithmic corrections do appear for infinitesimal
deviation from this point in the critical direction.

The second problem is that the extreme smoothness of a BKT transition, condition 3, makes
this kind of transition notoriously 
difficult to diagnose and locate accurately in numerical simulations. Here we use the fidelity susceptibility 
(FS) as a computationally simple and universally applicable indicator of a BKT transition. Large-scale 
numerical studies of BKT transitions by monitoring the FS have been successfully performed for
various one-dimensional quantum systems \cite{Sun2015,DamskiBose,Greschner}. 
But all the models considered previously display at least an explicit U(1) symmetry. There are
no FS studies of BKT transitions in quantum systems with discrete symmetries only. 
In addition, while \(p\)-clock models have been under investigation since the late 1970s, there seem to be 
no large-scale simulations of QPTs in  $p$-clock models for \(p>4\) available in the
literature. In this work we fill this gap. 

The success of the FS as a diagnostic tool of phase transitions in clock models is an important aspect 
of our present work. For example, there was a study of the $p=6$ quantum clock model with the method
of level spectroscopy\cite{Matsuo}
that relied on a Lanczos diagonalization of relatively short chains ($L=10$ or less sites).
This study is particularly relevant for our paper. In level-spectroscopy analyzes, \cite{Matsuo} 
twisted boundary conditions are imposed and the crossings between the $n$ and $n+1$ excited
 energy levels, with $n \sim p$, are identified and monitored. The BKT transitions that are associated with 
a local order parameter in the gapped phase (see Table\,\ref{MainTable}),  
can be located very accurately by these crossings \cite{Nomura}.
However, many transitions that are of the BKT-type by the full set of criteria 1 - 7 
do not  show any level crossings in the excited states. 
That is the case for example for Mott transitions in Bose-Hubbard 
chains at integer particle fillings and the BKT transitions occurring in the U(1) $p$-clock models introduced
and studied in this paper. In contrast, the FS
is an {\it unbiased} tool for diagnosing BKT transitions, meaning that this indicator always develops a 
distinct universal peak associated with the BKT transition regardless of the model.
This peak, caused by condition 7 (the square-root singularity in the correlation length $\xi$), 
appears within the gapped regime and shifts towards the BKT point with increasing system 
size. 

The paper is organized as follows. Section \ref{background} is mainly background material on the class 
of models and tools for identifying BKT transitions. In Section \ref{ZpU(1)models} we present the  long-range 
quantum self-dual $p$-clock models with a brief symmetry analysis. Section \ref{FidelityS} summarizes 
the key properties of the  FS, used in DMRG calculations  to identify BKT transitions.
Then, Section \ref{DMRGZp}, investigates the quantum phase diagram and nature of the phase transitions 
in \(p\)-clock models by means of symmetry  considerations and DMRG simulations. To understand the interplay between 
compactness of the clock degree of freedom and the emergence of a continuous symmetry, we introduce 
the  U(1) clock models and establish their quantum phase diagrams in Section \ref{U1clockmodel}. 
Section \ref{largep} studies the large-$p$ limit of the self-dual $p$-clock chain in the
low energy sector and derives, using exact methods, the $O(2)$ quantum rotor model. In particular, we determine 
the {\it exact} critical coupling of this latter model. Finally, in Section \ref{classicalmap}, we map the critical 
couplings of the quantum chains studied above to the critical temperatures 
$T_{\sf BKT}$ of their two-dimensional classical counterparts. Section \ref{Conclusion} summarizes the
main findings of our work.

\section{Background}
\label{background}


\subsection{$\mathbb{Z}_p$ long-range clock model}
\label{ZpU(1)models}

Clock models comprise a rich variety of natural 
generalizations of spin-\(1/2\) Hamiltonians. A clock degree
of freedom \cite{Schwinger,Ortiz} is described by a pair of unitary operators \(U, V\)
defined to extend the relation \(\sigma^x\sigma^z=e^{i2\pi/p}\sigma^z\sigma^x\), between 
Pauli matrices with 
\(p=2\), to arbitrary integers \(p>2\), 
\begin{equation}
\label{clock_edf}
VU=\omega UV,\,\,\,\,  U^p= V^p=\one,\,\,\,\,   \omega \equiv e^{i2\pi/p}. 
\end{equation}
In contrast, higher-spin generalizations of spin-\(1/2\) Hamiltonians 
emphasize the $su(2)$ Lie-algebraic properties of Pauli matrices. It follows from
Eq. \eqref{clock_edf} that the state space $\bar{\mathcal{H}}_{p}$ for a single clock 
degree of freedom
is \(p\)-dimensional, with orthonormal basis vectors
 \begin{equation}\label{pstate}
|s\rangle,\ \ \ \ s=0,\cdots,p-1 ,
\end{equation}
and, in that basis,  the operators \(U, V\) are represented irreducibly as
\begin{eqnarray} \hspace*{-0.5cm}
{U}=\begin{pmatrix}
1& 0& 0& \cdots& 0\\
0&\omega & 0& \cdots& 0\\
0& 0&\omega^2 & \cdots& 0\\
\vdots& \vdots& \vdots&      & \vdots \\
0& 0& 0& \cdots& 0\\
0& 0& 0& \cdots&\omega^{p-1} \\
\end{pmatrix} ,
{V}=
\begin{pmatrix}
0& 1& 0& \cdots& 0\\
0& 0& 1& \cdots& 0\\
0& 0& 0& \cdots& 0\\
\vdots& \vdots& \vdots&      & \vdots \\
0& 0& 0& \cdots& 1\\
1& 0& 0& \cdots& 0\\
\end{pmatrix},  
\label{VsandUs}
\end{eqnarray}
which generates the full algebra of ($p\times p$) complex matrices known as 
Weyl group algebra, where $V$ is known as the fundamental circulant matrix. 
These matrices reduce explicitly to \(\sigma^z, \sigma^x\) for \(p=2\). 

Clock degrees of freedom admit a kinematical interpretation first emphasized 
by H. Weyl, and later by J. Schwinger (see, e.g., [\onlinecite{Schwinger}]). Since
\begin{eqnarray}
U\ket{{s}}= \omega^{{s}} \ket{{s}} \ \mbox{ and } V\ket{{s}}= \ket{{s-1}}
\end{eqnarray}
\(V\) (\(V^\dagger\))
becomes the generator of counterclockwise (clockwise) unit jumps from any one
state to its previous (next) nearest neighbor (in the space of internal clock degrees 
of freedom). In this sense,
\(U, V\) represent conjugate position-like and momentum-like operators on the 
discretized circle. This interpretation is further
strengthened by the fact that the discrete Fourier transform, 
given in matrix form by 
\begin{eqnarray}
{F}^{\dagger}=\frac{1}{\sqrt{p}}
\begin{pmatrix}
1& 1& 1& \cdots& 1\\
1& \omega& \omega^2& \cdots& \omega^{p-1}\\
1& \omega^2& \omega^4& \cdots& \omega^{2(p-1)}\\
\vdots& \vdots& \vdots&      &\vdots \\
1& \omega^{p-1}& \omega^{(p-1)2}& \cdots& \omega^{(p-1)(p-1)}
\end{pmatrix} ,
\label{DFTmatrix}
\end{eqnarray}
satisfies 
\begin{equation}\label{wga_aut}
{F} U{F} ^\dagger=V^{\dagger},\qquad {F} V{F} ^\dagger=U ,
\end{equation}
and help determine the eigenvectors of $V$, 
$V\ket{\tilde{s}}= \omega^{\tilde{s}} \ket{\tilde{s}}$, 
via Fourier transformation of the eigenvectors of $U$ as
\begin{eqnarray}
 \ket{\tilde{s}} = \frac{1}{\sqrt{p}} \sum_{s=0}^{p-1}\omega^{\tilde{s} . s} 
 \ket{s} , \  \ \tilde{s}=0,1,\cdots,p-1  .
\label{eigenvofV1}
\end{eqnarray}
The \(U\) operator acts as a ladder operator in this basis since
\(U \ket{\tilde{s}}= \ket{\tilde{s}+1}\).
More  detailed discussion of clock degrees of freedom and 
reference to the original literature can be found in [\onlinecite{Ortiz}]. 

In the following, we will consider
clock variables $U_i, V_i, i=1,\dots,L$, arranged linearly, and commuting on different
sites of a chain of size $L$. The 
state space is simply the tensor product of single-clock Hilbert spaces $\bar{\mathcal{H}}_{p,i}$, i.e., 
\(\mathcal{H}_p=\bigotimes_{i=1}^L
\bar{\mathcal{H}}_{p,i}\).

A general type of quantum $p$-clock chain with simple duality properties is the {\it long-range} $p$-clock model
\begin{eqnarray}
H_p&=&H_{U}+H_V  ,
\label{Hpself-dual}
\end{eqnarray}
with
\begin{eqnarray}
H_{U}=-\frac{1}{2}\sum_{m=1}^{L-1}\sum_{i=1}^{L-m}
J_{i,m}(U_{i+m}^\dagger U^{\;}_i+ U_{i}^\dagger U^{\;}_{i+m}) ,
\label{Hpself-dual1}
\end{eqnarray}
and
\begin{equation}
H_{V}=-\sum_{i=1}^{L} h_i (V^{\;}_i+V_i^{\dagger}).
\end{equation}
This $L$-sites chain includes a kinetic-energy term written in terms of unitary operators $U_i$, 
and a potential energy term that involves unitary operators $V_i$. The corresponding coupling constants 
are $J_{i,m}$ and $h_i$, respectively. 

The most widely studied special instance \cite{book} is the quantum vector 
Potts model, with  
$J_{i,m}=J_x\delta_{i,1}$ and $h_i=h$
\begin{equation}
\label{TPCM}
H_{\sf VP}=-\frac{J_x}{2}\sum_{i=1}^{L-1} (U_{i+1}^\dagger U^{\;}_i+ U_{i}^\dagger U^{\;}_{i+1})+H_V,
\end{equation}
with
\begin{equation}
\label{H_V}
H_{V}=-h \sum_{i=1}^{L} (V^{\;}_i+V_i^{\dagger}),
\end{equation}
often referred to simply as {\it the} $p$-clock chain.
The vector Potts model is exactly self-dual and reduces to the transverse-field Ising chain for \(p=2\). 
Cases corresponding to the values of $p=3$ and $p=4$ are equivalent to the 3-states quantum 
Potts model, and two independent copies of transverse-field Ising chains \cite{Ortiz}, respectively.
Another instance of the long-range $p$-clock model studied before occurs for \(p=2\), $J_{i,1}=J,\ J_{i,2}=-\Delta$, 
\(J_{i,m}=0,\ m >2\), and \(h_i=h\).  
In this case, the Hamiltonian \(H_p\) of Eq. \eqref{Hpself-dual}
corresponds to the ANNNI model\cite{ANNNIbook},
\begin{equation}\hspace*{-0.15cm}
H_{\sf ANNNI}=-J\sum_{i=1}^{L-1} 
\sigma^z_{i}\sigma^z_{i+1}+ \Delta\sum_{i=1}^{L-2}\sigma^z_i\sigma^z_{i+2}
-h\sum_{i=1}^L \sigma^x_i . 
\end{equation}

The ANNNI model \cite{ANNNI, ANNNIbook} shows a BKT transition with all 1 - 7 conditions satisfied. In particular, 
the emergence of a U(1) symmetry can be understood in the limit of $h=\Delta \to \infty$, when the model 
becomes equivalent to two decoupled transverse-field Ising chains at the critical value of their 
transverse field \cite{Allen} (described by two copies of massless Majoranas 
equivalent to a single Dirac fermion with U(1) particle-number conservation).
We call the corresponding clock generalization the \(p\)ANNNI model. The
\(p\)ANNNI model shows some combination of the remarkable
behaviors of the vector Potts\cite{Ortiz} (for \(p\geq 5\)) and 
ANNNI models \cite{ANNNIbook}, in particular a plethora of commensurate and incommensurate phases. 
Reference\,[\onlinecite{Milsted14}] investigated the $p$ANNNI model for  \(p=6\).

As shown in Ref.  [\onlinecite{Ortiz}], the group of symmetries of the 
standard $p$-clock model is 
non-Abelian for $p\geq 3$. Similarly, for the long-range $p$-clock model, 
consider the two Hermitian operators
\begin{equation}\label{polyhedral}
\mathcal{C}_0=\prod_{i=1}^L C_{0 i},\ \ \ \ \ \ 
\mathcal{C}_1=\prod_{i=1}^L C_{1 i},
\end{equation}
whose action on the basis which spans ${\cal H}_p$ is 
\begin{equation}
C_{0 i}|s_i\rangle=|-s_i\rangle,\ \ \ \ C_{1 i}|s_i\rangle=|1-s_i\rangle,
\end{equation}
where modular, mod($p$), arithmetic is assumed. The operator 
$\mathcal{C}_0$ is the exact analogue of the
parity operator $\mathcal{P}$ for position eigenstates on the real line, 
$\mathcal{P}|x\rangle=|-x\rangle$, 
and it is related  to the discrete Fourier transform,  ${\cal F} = \prod_{i=1}^L F_i$,  as 
${\cal F}^2=({\cal F}^\dagger)^2={\cal C}_{0}$. Notice that 
$C_{0i}C_{1i}=V_i$.

The action of $\mathcal{C}_0$ and $\mathcal{C}_1$ on the
discrete position \(U_i\) and momentum \(V_i\) operators is
\begin{eqnarray}
\mathcal{C}_0V_i\mathcal{C}_0&=&V_i^\dagger,\ \ \ \ \ \ \ \ \ \  
\mathcal{C}_1V_i\mathcal{C}_1= V_i^\dagger, \\
\mathcal{C}_0U_i\mathcal{C}_0&=&U_i^\dagger,\ \ \ \ \ \ \ \ \ \ 
\mathcal{C}_1U_i\mathcal{C}_1=\omega U_i^\dagger.
\end{eqnarray}
One can then show that the operators 
$\mathcal{C}_0$ and $\mathcal{C}_1$ commute with the Hamiltonian 
$H_p$ and satisfy 
\begin{equation}\label{polyhedral1}
\mathcal{C}_0^2=\mathcal{C}_1^2=(\mathcal{C}_0\mathcal{C}_1)^p={\one},
\end{equation}
indicating that they are not only Hermitian but also unitary operators.
The product, $\mathcal{C}_0\mathcal{C}_1\equiv\hat{Q}$ 
known as the $\mathbb{Z}_p$ charge,  is the standard Abelian symmetry  
of the $p$-clock model $H_{\sf VP}$. It is the symmetry that breaks down
spontaneously in the low-temperature ordered phase \cite{Ortiz}.

The Hamiltonian $H_p$, for $p$ even and $J_{i,m}=0$ with $m$ even, 
has a spectrum that is symmetric 
about zero. In other words, one can construct a unitary operator
\begin{eqnarray}
{\cal U}_c=\left ( \prod_{i=1}^L  (U_i)^{\frac{p}{2}}\right )\left ( \prod_{i=1}^{[L/2]}  
(V_{2i})^{\frac{p}{2}}\right ) ,
\end{eqnarray}
that anticommutes with $H_p$, i.e., ${\cal U}_c \, H_p\, 
{\cal U}^\dagger_c=-H_p$. In particular, the transverse-field 
Ising and ANNNI models share this particle-hole symmetry. 
The spectrum is not symmetric about zero when 
$p$ is odd.

\subsection{A smoking gun of BKT Transitions:\\ Peaks in the Fidelity Susceptibilty}
\label{FidelityS}

For one-dimensional quantum many-body systems that experience a QPT in the BKT 
universality class, condition 3 must hold. Let $H(\lambda)$ be the
Hamiltonian of the system, dependent on some control parameter $\lambda$.
Then,  $\varepsilon_0(\lambda)= \langle \psi_0(\lambda)| H(\lambda)| \psi_0 (\lambda)\rangle/L$ 
remains a smooth function of $\lambda$, that is, differentiable to any finite order, even at the BKT point  
$\lambda=\lambda_{\sf BKT}$, and in the thermodynamic limit. Nonetheless,   $\varepsilon_0(\lambda)$ 
develops an essential singularity at $\lambda_{\sf BKT}$ and the gap in the excitation spectrum 
opens exponentially slow when departing from the critical point. Thus, it is difficult to locate the 
BKT transition by following the behavior of the energy gap directly or by way of the correlation length. 

To study and identify 
QPTs, in this work we use the method based on the ground state FS \cite{Zanardi,You,Gu10rev}.
It turns out that one can detect a BKT transition by following numerically the behavior of the overlap of 
two infinitesimally close 
ground state wave-functions \cite{Sun2015}. The FS, defined as
\begin{equation}
\label{FS}
\chi_L(\lambda)=-\frac{2}{L}\lim_{\delta \to 0} \frac{ \ln | \langle   \psi_0 (\lambda) | \psi_0 (\lambda+\delta)  \rangle|}{\delta^2},
\end{equation}
is straightforwardly calculated in any computational method, especially  large-scale DMRG 
\cite{White}, that determines  ground states efficiently. In addition, the FS is a powerful tool
for computing response functions such as the dynamical conductivity of interacting electrons
that are notoriously hard to calculate by other methods\cite{Greschner2013}. Although the FS cannot be 
measured directly in experiments, it could be measured indirectly by way of these response functions \cite{GuYu2014}.


It has been established rigorously in recent works \cite{Sun2015,Vekua2016} 
that the FS, $\chi_L(\lambda)$, stays finite across a BKT transition even in the thermodynamic 
limit $L\to \infty$. Nevertheless, within a low-energy effective-field-theory approach, 
it was shown \cite{Sun2015} that the FS develops a characteristic peak near $\lambda_{\sf BKT}$.
In the same work \cite{Sun2015}, the scaling with system size of the height and position of the peak  were determined. 
The peak height approaches its infinite-system-size limit as 
\begin{equation}
\label{heights}
\chi_{\infty} -\chi_L(\lambda^m) \sim \frac{1}{ \ln{(bL)}}   + \cdots ,
\end{equation}
 and the peak position always moves from the gapped phase towards the BKT point as $\sim 1/ \ln^{2}{(cL)}$.
 The constants $b$ and $c$ are positive, non-universal numerical coefficients, the dots indicate 
 sub-leading contributions, and $\lambda^m=\lambda^m(L)$ is the value of the coupling constant 
 where the finite-size peak is located.  The fact  that the position of the peak moves from the gapped 
 region towards the BKT transition point $\lambda_{\sf BKT}$ with increasing system size is a 
 particularly attractive feature for disentangling two nearby-located BKT transition points in finite-size 
 studies when the critical phase is narrow. 

We conclude this discussion by pointing out the interplay between the FS and 
self-duality transformations. In the following we will use the formula 
\begin{eqnarray}\label{diff_fs}
\chi_L(\lambda)=-\frac{1}{L}\mbox{Re}\langle \psi_0(\lambda)|\frac{d^2}{d\lambda^2 }|\psi_0(\lambda)\rangle.
\end{eqnarray}
for the FS. A large class of self-dual Hamiltonians, including the quantum \(p\)-clock chains, are of the form
\(H(\lambda)=A+\lambda B.\) The self-duality transformation of these models
is a unitary transformation\cite{dualitiesPRL} \(U_{\sf d}\) that exchanges \(A\) and \(B\).
As a consequence,
\begin{eqnarray}
\label{sd}
U_{\sf d}H(\lambda)U_{\sf d}^\dagger=\frac{1}{\lambda}H(\frac{1}{\lambda}).
\end{eqnarray}
We will assume that \(\lambda>0\) and the ground state unique. For many models,
one can arrange to have an exact self-duality transformation and a unique ground state 
by adjusting the boundary conditions\cite{AdvPhys}. Then, 
\[|\psi_0(\frac{1}{\lambda})\rangle=U_{\sf d}|\psi_0(\lambda)\rangle\] and, 
since \(U_{\sf d}\) is independent of \(\lambda\), it follows from Eq.\,\eqref{diff_fs} that
\[
\chi_{L}(\lambda)=\frac{1}{\lambda^4}\chi_L(\frac{1}{\lambda}).
\] 
This equation gives an exact relation between pairs of peaks of the FS at positions \(\lambda_-\lambda_+=1\) 
related by self-duality, but only in limit \(\delta \rightarrow 0\) of Eq.\,\eqref{FS} and for appropriate self-dual  
boundary conditions.  

\begin{figure}[htb]
\includegraphics[width=8cm]{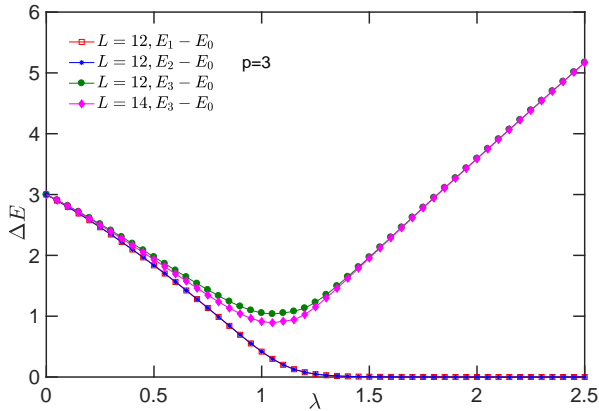}
\caption{Energy gaps for low-energy levels as a function of $\lambda$ for $p$=3
and measured with respect to the ground energy level \(E_0\). For $\lambda> 1$ the 
ground state is 
$p$-fold degenerate in the thermodynamic limit. For $\lambda=0$ the ground state is 
unique. However, the first excited state is $L$-fold degenerate. A qualitatively 
similar picture is obtained for the $p=2$ and $p=4$ cases as well, with a well pronounced 
single minimum in the excited-states energy 
gap near $\lambda=1$.}
\label{fig:levelsp03}
\end{figure}

\section{Quantum $\mathbb{Z}_p$ Clock Model}
\label{DMRGZp}

In this section we present some of the main results of our work: the study of QPTs in 
quantum $p$-clock chains (see Eq. (\ref{TPCM})) for $p\ge 5$. Our findings are  based on
a detailed investigation of the FS of these models. The FS turns out to be an 
invaluable tool to unveil its quantum critical behavior. At $\lambda=0$ one can obtain 
an analytic expression for the FS. At $L=\infty$ and for any $p$, 
by performing a second-order perturbation theory calculation, one gets 
\begin{equation}
\label{er}
\chi_{\infty}(0)=\frac{(1+\delta_{p,2})}{8(1-\cos{(2\pi/p)})^2}.
\end{equation}
One can see from this formula that at $\lambda=0$ and for $p\gg 1$, the  FS 
behaves as $\chi_{\infty}(0)\sim p^2$. For generic values of $p$ and $\lambda$ we will
compute the FS numerically. 

\begin{figure}[htb]
\includegraphics[width=8cm]{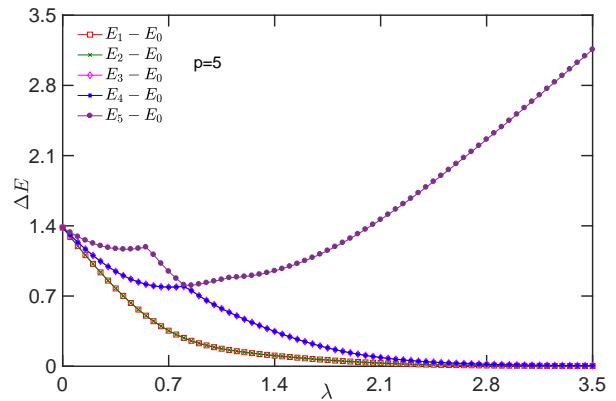}
\caption{Energy gaps for $p=5$ and $L=8$ showing clearly a more complex structure
than the $p=3$ case shown in Fig. \ref{fig:levelsp03}. There are a few 
level-crossings between excited states. For large 
$\lambda$, the ground state becomes $5$-fold degenerate.}
\label{fig:EL03}
\end{figure}

\subsection{The low-energy spectra}

It is instructive to begin with the behavior of the low-energy levels as a function of 
the ratio $\lambda=J/(2h)$, for different values of $p$. For 
$p\le 4$, the quantum $p$-clock chain supports one critical (gapless) point  separating 
two gapped phases at the self-dual point $\lambda=1$\cite{Ortiz}: A disordered phase is 
realized for $\lambda\le 1$, and an ordered 
phase with spontaneously broken $\mathbb{Z}_p$ symmetry is realized for $\lambda>1$. The value 
of the central charge $c$ (a measure of the number of degrees of freedom associated to gapless modes) is 
$C=1/2$ for $p=2$, $c=4/5$ for $p=3$, and $c=1/2+1/2=1$ for $p=4$. In Fig. \ref{fig:levelsp03}, as a typical 
representative of the $p\le 4$ cases, we show the behavior of the lowest energy levels (with
the ground state energy $E_0$ subtracted) for $p=3$ as a function of $\lambda$ for a rather small system 
size ($L=12$ and for the true excited state $L=14$ as well). The energy levels were computed by 
exact diagonalization. As one can observe in Fig. \ref{fig:levelsp03}, 
for $\lambda<1$ the ground state is unique (for $\lambda=0$, when the $p$-clock chain Hamiltonian 
becomes site-decoupled, the first excited state is $L$-fold degenerate). On the other hand, for 
$\lambda > 1$, the three-fold degeneracy of the ground state is evident. The true excited state 
(indicated by $E_3-E_0$ in Fig. \ref{fig:levelsp03}) has a minimum excitation gap (a finite-size gap) 
at $\lambda \simeq 1$. That  gap decreases with increasing $L$. The low-energy level 
structure for $p=2$ and $p=4$ is similar.

However, starting from $p=5$ and for all $p>5$ cases the diagrams of the lowest-energy levels 
become more complex. In Fig. \ref{fig:EL03}, as a typical representative of the $p\ge 5$ cases, we present 
the behavior of the lowest-energy gaps for $p=5$. One can see that the behavior of 
energy levels are now more involved than for the $p\le 4$ cases, and namely we do not see a well 
pronounced unique minimum in the excited states as before, instead there is a complex structure with 
several level crossings in the excited spectra for $0.5<\lambda<1.2$. For $p=6$ the above mentioned 
level-crossings in the excited states were used to locate the BKT phase transitions \cite{Matsuo}. 
We will  instead 
rely on the FS to study phase transitions in quantum $p$-clock chains.

\subsection{The FS of quantum $p$-clock chains for $p\le 4$}

At a quantum critical point, the singular (diverging in thermodynamic limit) behavior of the FS peak height obeys the following scaling law 
\begin{equation}
\label{SSR}
\chi_L(\lambda^m) \sim  L^{\frac{2}{\nu}-1} \sim L^{2d+2z-2\Delta-1},
\end{equation}
where $\nu$ is the critical exponent for the correlation length. In $p$-clock models we consider the dimensionality 
of space $d$ and the dynamical critical exponent $z$ to be both equal to unity. The dimension of the 
operator driving the phase transition and the value of $\nu$ at criticality are \cite{DKMM94,CFT} $\Delta=\nu=1$ for $p=2$,
 and $\Delta=4/5$, $\nu=5/6$,  for $p=3$. Hence, for $p=2$ and $p=3$, the FS at criticality ($\lambda=1$)
is divergent in the thermodynamic limit. This divergence should be traceable as a peak for 
finite $L$ with the height 
of the peak increasing as $\sim L$ for the Ising case \cite{CWHW,DamskiIsing,Azimi,USIsing,Gaoyong} 
and as $\sim L^{7/5}$ for the $3$-states Potts model. 
\begin{figure}
\includegraphics[width=8cm]{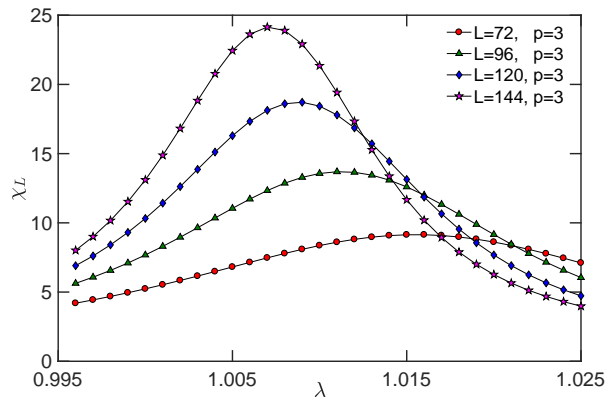}
\caption{ FS $\chi_L$ for $p=3$ and system sizes $L=72,96,120$, and $144$. One can observe clearly
a single peak in the FS. The peak position moves towards the self-dual point $\lambda=1$ and the height 
of the peak increases with increasing system size. Similarly, for $p=2$ and $p=4$ there is a 
single peak moving towards the self-dual point $\lambda=1$ with increasing system size.}
\label{fig:levelsp3}
\end{figure}
In Fig. \ref{fig:levelsp3}, as a typical representative of the $p\le 4$ cases, we present our numerical data 
of the FS for $p=3$. One can observe a single peak in $\chi_L$ in the vicinity of $\lambda=1$, moving 
towards the self-dual point $\lambda=1$ with increasing system size. 
 
\begin{figure}[ht]
\includegraphics[width=8cm]{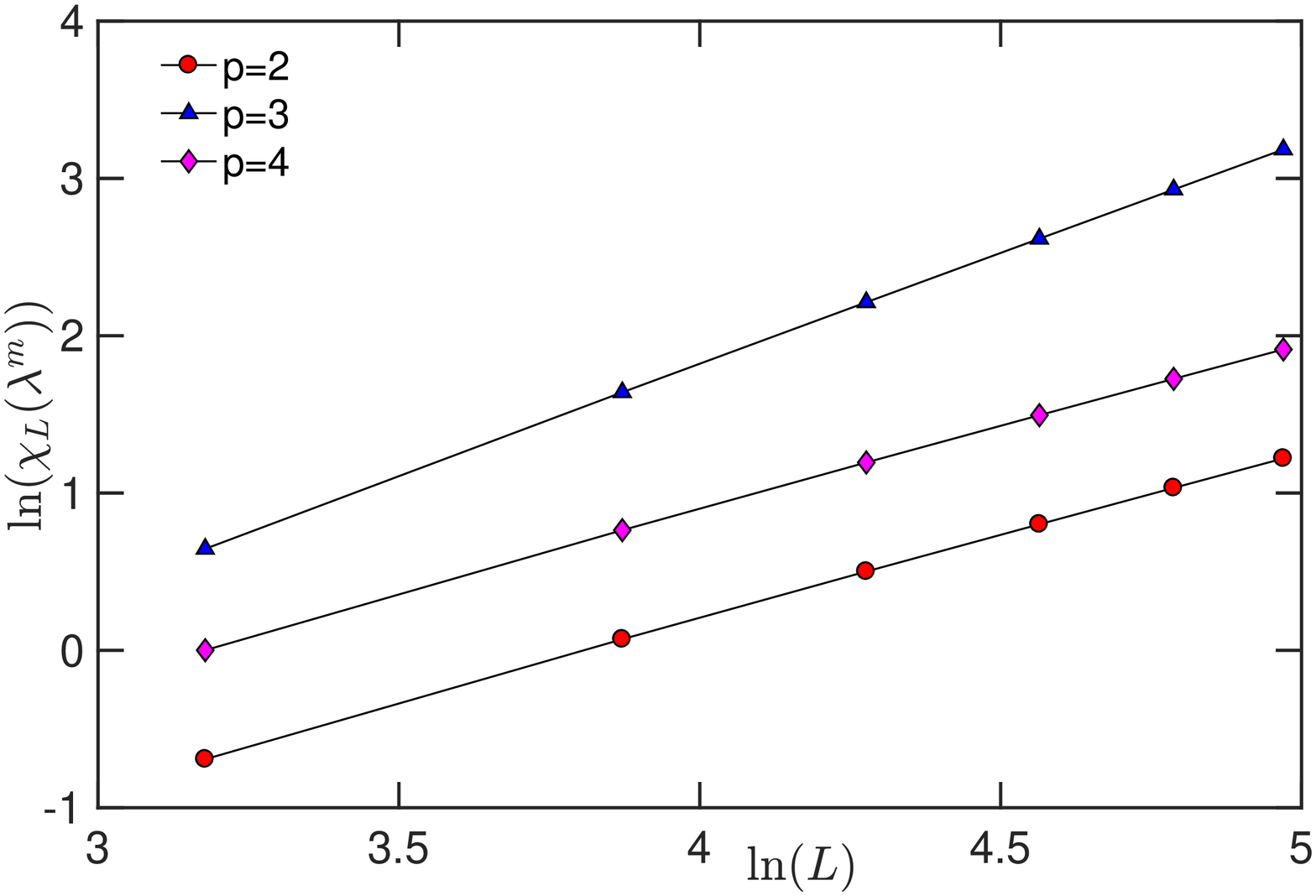}
\caption{Log-log plot of the FS peak height versus system size $L$ for quantum $p$-clock 
chains with $p=2,3$ and $4$. The height of the peaks increases as $\sim L$ (two parallel continuous lines connecting 
little circles and diamonds) for $p=2$ and $p=4$, while it scales as $\sim L^{1.4}$ (continuous line connecting 
little triangles) for $p=3$. These results are in perfect agreement with Eq. (\ref{SSR}).}
\label{fig:p2p3}
\end{figure}

The quantum $p$-clock chain for $p=4$, as already mentioned, is dual to two decoupled quantum Ising 
chains \cite{Ortiz} and at criticality its symmetry is enhanced to U(1). Hence, the overlap of ground states at 
two different couplings  is related to the overlap of ground states of quantum Ising chains as 
$| \langle   \psi^{p=4}_0 (\lambda) | \psi^{p=4}_0 (\lambda+\delta)  \rangle|= | \langle   
\psi^{p=2}_0 (\lambda) | \psi^{p=2}_0 (\lambda+\delta)  \rangle|^2$ and thus, from the definition, Eq. (\ref{FS}), it follows that
\begin{equation}
\label{ATI}
\chi^{p=4}_L(\lambda)=2 \chi^{p=2}_L(\lambda).
\end{equation}
In Fig. \ref{fig:p2p3} we present a log-log plot of the scaling of the FS peak height with system size 
for $p=2,4$ and $p=3$. Our numerical simulations are in excellent agreement with the results of 
Eq. (\ref{SSR}) based on simple scaling arguments.

\subsection{The FS of quantum $p$-clock chains with $p \ge 5$}
\label{FSpgeq5}

For $p\ge 5$, according to analytical predictions \cite{Elitzur, Ortiz}, we expect two BKT transition points. The dimension of the operator 
driving the phase transition is\cite{GiamarchiBook} $\Delta=2$ and $d=z=1$ at both critical points. 
Since $\Delta>3/2$, the only relevant information that Eq. (\ref{SSR}) provides is that FS peak height does not diverge when taking the thermodynamic limit at the BKT transition, that is, the phase transition is not {\it sharp}. 
Nevertheless, the BKT phase transitions should be detectable by corresponding peaks in the FS due to arguments 
that go beyond the simple scaling reasonings that lead to the estimate in Eq. (\ref{SSR}).

\begin{figure}[htb]
\includegraphics[width=8cm]{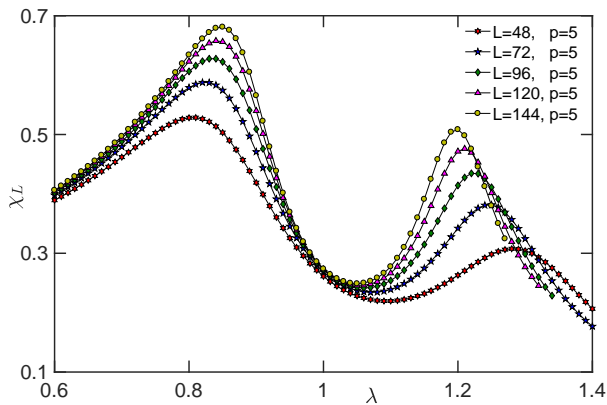}
\caption{ FS $\chi_L$ for $p=5$ and system sizes $L\le 144$. A double hump structure in the FS curves 
is clearly visible, with peaks becoming narrower and higher with increasing system size. The peak 
positions $\lambda_-$ and $\lambda_+$ shift towards each other. They  move from the gapped 
regions towards their thermodynamic values $\lambda^*_{-}$ and $\lambda^*_{+}$ at the boundaries of the critical phase, with increasing system size.}
\label{fig:pd5}
\end{figure}
\begin{figure}
\includegraphics[width=8.5cm]{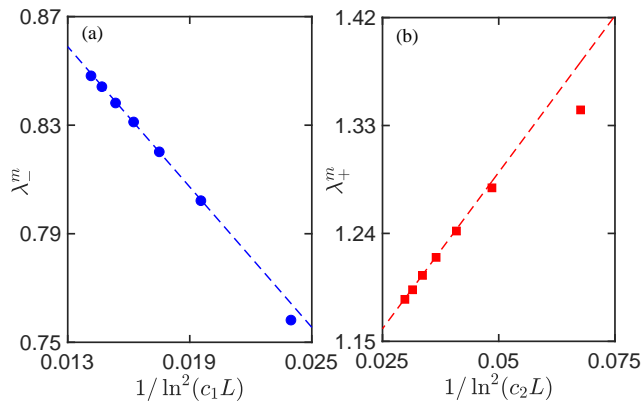}
\caption{Scaling of the FS peak positions with system size for $p=5$ and system sizes of up to 
$L=168$ sites. The numerical constants $c_1$ and $c_2$ are positive. The extrapolated values of the 
peak positions are (a) $\lambda^*_{-}\simeq 0.966$ and (b) $\lambda^*_{+} \simeq 1.035$. For large 
$L$, the fits follow accurately the BKT scaling law Eq. (\ref{heights}), 
confirming the BKT nature of the underlying QPTs.}
\label{fig:pd50}
\end{figure}

The behavior of the FS for $p=5$, and system sizes $L=48,72,96,120$ and $144$, is presented in 
Fig. \ref{fig:pd5}, where one can clearly see a double-hump structure in the FS curves, in contrast to a 
single (and sharper) peak observed for $p \le 4$. As shown in Fig. \ref{fig:pd50}, the peak 
locations $\lambda^m_{-}<1$ and $\lambda^m_{+}>1$
shift from the gapped phases towards each other with increasing $L$ in excellent agreement with 
the BKT scaling law of Eq. (\ref{heights}). The extrapolated values yield an accurate estimate of the width of the critical phase 
which is at least as narrow as \(\lambda^*_{+}-\lambda^*_{-}=0.079\).

\begin{figure}[htb]
\includegraphics[width=8cm]{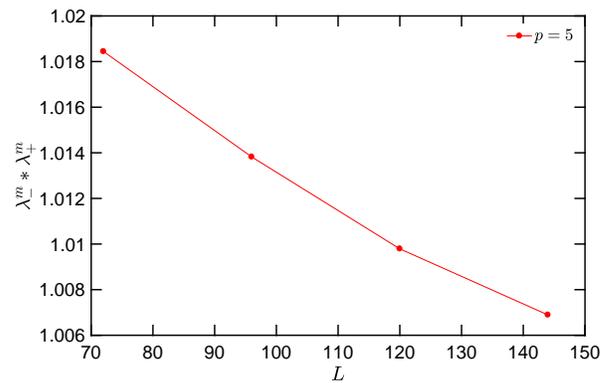}
\caption{Evolution of the product of the two peak positions $\lambda_{-}^m \lambda_{+}^m$ towards 
unity with increasing system size for $p=5$.}
\label{fig:ProductEvolution}
\end{figure}

The product of the two peak positions $\lambda_{-}^m \lambda_{+}^m$ 
approaches unity to a very good accuracy with increasing $L$ as one can see in 
Fig. {\ref{fig:ProductEvolution}}. The product of the extrapolated values 
(see Fig.\,\ref{fig:pd50}) is \(\lambda_-^*\lambda_+^*=0.99981\), that is, exactly unity within our
precision. The Hamiltonian given 
in Eq. (\ref{TPCM}) displays an approximate self-duality relation \cite{Ortiz} that
becomes asymptotically exact in the thermodynamic limit. Actually, an exact self-duality also exists for finite chains and  
special types of open boundary conditions \cite{AdvPhys, Ortiz}. Due to this 
self-duality symmetry, if there is a single critical point it must lie at the self-dual point $\lambda=1$ in 
thermodynamic limit, as is the case for $p\le 4$. If there are two critical points,
then they must occur {\it symmetrically} with respect to the  self-dual point $\lambda=1$ and, 
moreover, both critical points must belong to the same universality class \cite{Ortiz}. 
In particular, the self-dual relation  implies that $\lambda^*_-\lambda^*_+=1$. 
Since we simulate Eq. (\ref{TPCM}) which does not 
show the exact self-duality relation of Eq.\,\eqref{sd}, we only see the self-duality
symmetry emerge to a very good approximation with large system size. 

\begin{figure}[ht]
\includegraphics[width=8cm]{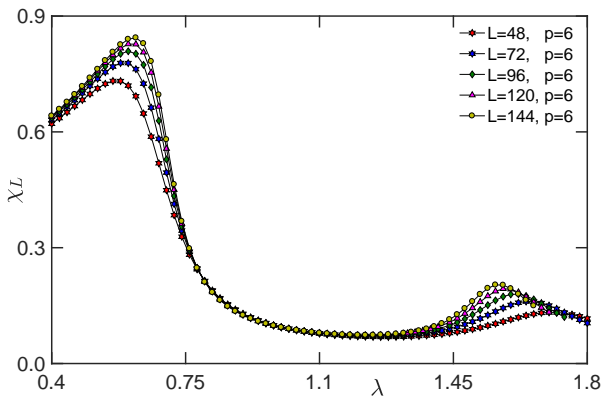}
\caption{The FS $\chi_L$ for $p=6$ and system sizes $L\le 144$. The peak positions are further away from 
each other when compared to the $p=5$ case shown in Fig. \ref{fig:pd5}. Similar double-hump structure 
in the FS is observed for other values of $p>6$ as well, with peak positions even 
further away from each other. The gapless parameter region is expected to increases as $p$ increases\cite{foot}.}
\label{fig:pd6}
\end{figure}

\begin{figure}[hbt]
\includegraphics[width=8.5cm]{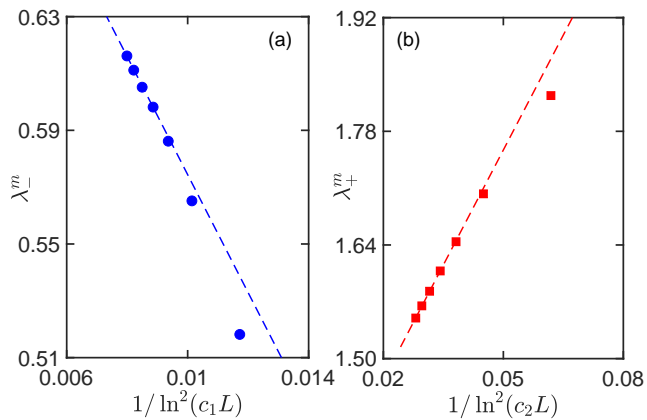}
\caption{Scaling of the FS peak positions with system size for $p=6$ and system sizes 
up to $L=168$ sites; $c_1$ and $c_2$ are positive numerical constants. The extrapolated 
values of the peak positions are (a) $\lambda^*_{-}\simeq 0.782$ and (b) 
$\lambda^*_{+} \simeq 1.285$ (\(\lambda_-^*\lambda_+^*=1.005\)) and are consistent
with the transition points obtained by the level-spectroscopy method \cite{Matsuo}.}
\label{fig:pd60}
\end{figure}

\begin{figure}
\includegraphics[width=9cm]{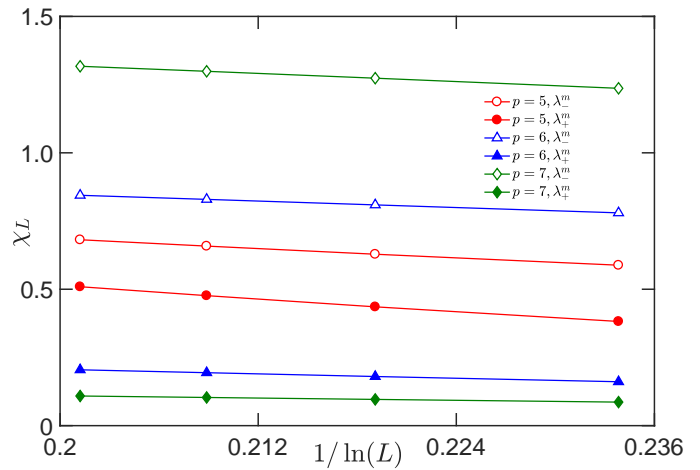}
\caption{Scaling of the FS peak heights with system size for $p=5,6$ and $7$. For the system 
sizes shown, $(L\ge 72)$, the peak heights follow, in excellent agreement, the leading 
behavior in Eq. (\ref{heights}), thus confirming the BKT nature of the QPTs.}
\label{fig:pd7}
\end{figure}

The behavior of the FS for $p=6$ is shown in Fig. \ref{fig:pd6} and the corresponding scaling 
of the 
peak positions are presented in Fig. \ref{fig:pd60}. One notes that, as compared to the 
$p=5$ case  in Fig. \ref{fig:pd5}, the peak positions are further apart for $p=6$.
This observation is consistent with the expectation \cite{foot} that the gapless parameter regime should
broaden with increasing $p$. For \(p=6\), our best estimate for the width of the critical 
phase is \(\lambda_+^*-\lambda_-^*=0.503\), see Fig.\,\ref{fig:pd60}.  
The FS peak heights for $p\ge 5$ increase more slowly with system size than for the cases 
$p\le 4$. Namely, as shown in Fig. \ref{fig:pd7}, for $p=5, 6$ and $7$, the peak heights converge to
infinite-system-size values logarithmically, in agreement with the BKT scaling law of Eq. (\ref{heights}). 

In summary, our method provides unambiguous numerical proof of the existence of two consecutive BKT phase 
transitions in the self-dual quantum $p$-clock chain for  $p=5, 6, 7$. Unfortunately, it is very difficult to extrapolate 
accurately the locations of the FS peaks in the thermodynamic limit for $p>5$. 
One encounters two problems with increasing $p$. First, the dimension of the local Hilbert 
space increases, since there is no conserved U(1) quantum number to take advantage of. The second
related and more severe problem is that to reach convergence for the position of the FS peaks one 
would need to reach larger system sizes than we are able to. As it turns out, for $p=5$ we have reached 
good convergence as evinced in Fig. \ref{fig:pd50}. For $p=6$ we profited from the highly precise values 
of the BKT transition points obtained by the level-spectroscopy method \cite{Matsuo} to fit our curves in 
Fig. \ref{fig:pd60}. For \(p=7\) the values for the FS peak {\it positions} have not converged with system size,
meaning that the variance in the available data points is too large. For this reason, we cannot provide
a reliable estimate of the transition points \(\lambda^*_\pm\) for \(p=7\). In contrast, and surprisingly, the
data for the {\it height} of the FS peaks do appear stable and extrapolate well to the thermodynamic limit. 
This is why we are able to exploit the BKT scaling law Eq. (\ref{heights}) to ascertain that the transitions for
\(p=7\) are BKT even as we cannot quite locate them to a high accuracy, see Fig.\,\ref{fig:pd7}. 
A similar situation is observed for $p=8$ (data not shown).

\section{Phase Transitions in the Quantum U(1) Clock Model}
\label{U1clockmodel}

In this section we address the role of symmetries on the quantum phase diagram of 
quantum $p$-clock chains. The standard $\mathbb{Z}_p$ clock chains of previous sections
enjoyed only discrete symmetries and a self-duality transformation, resulting in 
an ordered $p$-fold degenerate ground state in certain parameter regimes and a dual 
disordered phase. Below we introduce an explicitly U(1)-symmetric minimal modification 
of the quantum $p$-clock chain. With our techniques, several  U(1)-symmetric models 
can be constructed as ``minimal" modifications of the \(p\)-clock chain. Most
of these models satisfy conditions 1 and 2 for some parameter regime, and none of them
is self-dual. However, our  investigation numerically shows that the
model we introduce next retains at least one BKT transition,  for $p\ge 5$. For that reason we call it {\it the} 
U(1)-symmetric quantum \(p\)-clock chain, or simply the U(1) clock model for short.

\subsection{The U(1)-symmetric quantum \(p\)-clock chain}
\label{defU1clock}

Consider next a unitarily equivalent representation of the Weyl algebra. 
This new representation differentiates the odd from the even $p$ cases. Define the unitary transformation
\begin{eqnarray}
{\cal U}=\begin{cases}
V^{\frac{p-1}{2}} \ , p \in {\sf odd}  \\
V^{\frac{p-2}{2}} \ , p \in {\sf even} 
\end{cases} ,
\end{eqnarray}
and let us conjugate the standard representation to obtain
\begin{eqnarray}
{\cal U}^\dagger U  {\cal U}=\tilde{U} \  , \
{\cal U}^\dagger V  {\cal U}=V \ ,  \ {\cal U}^\dagger F {\cal U}=\tilde{F} ,
\end{eqnarray}
where, for $p \in$ odd,
\begin{eqnarray} \hspace*{-0.5cm}
\tilde{U}=\mbox{\sf diag}(\omega^{\frac{p+1}{2}},\omega^{\frac{p+3}{2}},\cdots,1,\cdots,
\omega^{\frac{p-3}{2}} , \omega^{\frac{p-1}{2}} ) ,
\end{eqnarray}
while for $p \in$ even
\begin{eqnarray} \hspace*{-0.2cm}
\tilde{U}=\mbox{\sf diag}(\omega^{\frac{p}{2}+1},\omega^{\frac{p}{2}+2},\cdots,1,\omega,\cdots,
\omega^{\frac{p}{2}-1} , \omega^{\frac{p}{2}} ) .
\end{eqnarray}
Notice that  
\begin{equation}
\tilde{F} \tilde{U} \tilde{F} ^\dagger=V^{\dagger},\qquad \tilde{F} V \tilde{F} ^\dagger=\tilde{U} .
\end{equation}

Given the generators of the SU(2) algebra in the spin $S=(p-1)/2$ representation
\begin{eqnarray} 
{S^z}&=&
\begin{pmatrix}
\frac{p-1}{2}& 0& 0& \cdots& 0 & 0\\
0& \frac{p-3}{2}& 0& \cdots& 0 & 0\\
\vdots& \vdots& \vdots&      &\vdots  & \vdots \\
0& 0& 0& \cdots&\frac{3-p}{2} & 0  \\
0& 0& 0& \cdots & 0 & \frac{1-p}{2}\\
\end{pmatrix}, \nonumber \\
S^+&=&\begin{pmatrix}
0& {\scriptstyle \sqrt{p-1}}& 0& \cdots& 0 & 0\\
0& 0& {\scriptstyle \sqrt{2(p-2)}}& \cdots& 0 & 0\\
\vdots& \vdots& \vdots&      & \vdots & \vdots \\
0& 0& 0& \cdots& {\scriptstyle \sqrt{2(p-2)}}& 0  \\
0& 0& 0& \cdots & 0 & {\scriptstyle \sqrt{p-1}}\\
0& 0& 0& \cdots & 0 & 0\\
\end{pmatrix} , \nonumber
\label{VsandUsandS}
\end{eqnarray}
and $S^-=(S^+)^\dagger$, one may study the transformation properties of 
the Weyl's group generators $\tilde{U}$ and $V$ under the U(1) transformation
\begin{eqnarray}
\label{symg} 
\mathcal{U}_\phi=e^{-\im \phi S^z},
\end{eqnarray} 
which commutes with $\tilde{U}$. To obtain the transformation of $V$ let us 
rewrite $V$, Eq. \eqref{VsandUs}, as the sum of two 
operators $V=\hat{V} + \hat{\Delta} $ with 
\begin{eqnarray} 
\hat{\Delta}=
\begin{pmatrix}
0& 0& 0& \cdots& 0\\
0& 0&0& \cdots& 0\\
\vdots& \vdots& \vdots&      & \vdots \\
0& 0& 0& \cdots& 0\\
1& 0& 0& \cdots& 0\\
\end{pmatrix}
=\frac{1}{(p-1)!} (S^-)^{p-1}\ , 
\end{eqnarray} 
i.e., the matrix that has only a $1$ in the lower-left corner. Then,
\begin{eqnarray}
\label{symcjh}
\mathcal{U}_\phi   \hat{V} \, \mathcal{U}_\phi^\dagger = e^{-\im \phi}
\hat{V} \ ,  \mbox{ and } \mathcal{U}_\phi  \, \hat{\Delta} \,
\mathcal{U}_\phi^\dagger  = e^{\im (p-1) \phi} \hat{\Delta} .
\end{eqnarray}

One can define a unitarily equivalent long-range $p$-clock model by replacing $H_U$ by 
$H_{\tilde{U}}$ in Eq. \eqref{Hpself-dual}, which for the standard $p$-clock chain becomes 
\begin{equation}
H_{\sf VP}=-\frac{J_x}{2}\sum_{i=1}^{L-1}(\tilde{U}_{i+1}^\dagger \tilde{U}^{\;}_i+ 
\tilde{U}_{i}^\dagger \tilde{U}^{\;}_{i+1})+H_V,
\end{equation}
with Fourier transformed representation
\begin{eqnarray}
\label{TPCMF}
H_{\sf VP}&=&-\frac{J_x}{2}\sum_{i=1}^{L-1} ({V}_{i+1}^\dagger {V}^{\;}_i+
 {V}_{i}^\dagger {V}^{\;}_{i+1}) \nonumber \\
&& -h\sum_{i=1}^{L} (\tilde{U}^{\;}_i+\tilde{U}_i^{\dagger}).
\end{eqnarray}

The transformation properties of Eq. \eqref{symcjh} motivate us to introduce the U(1) symmetric 
 $p$-clock model
\begin{eqnarray}
\label{HU1}
H_{\rm U(1)}&=&-\frac{J_x}{2}\sum_{i=1}^{L-1} (\hat{V}_{i+1}^\dagger \hat{V}^{\;}_i+ 
\hat{V}_{i}^\dagger \hat{V}^{\;}_{i+1}) \nonumber \\
&& -h\sum_{i=1}^{L} (\tilde{U}_i+\tilde{U}_i^{\dagger}),
\end{eqnarray}
where $[H_{\rm U(1)}, e^{-\im \phi S_T^z}]=0$, with $S_T^z=\sum_{i=1}^L S_i^z$.
In Ref. [\onlinecite{Ortiz}] the U(1) symmetry generated by $\exp[-\im \phi S_T^z]$ was 
identified as emergent for the $\mathbb{Z}_p$ clock model.

For $p=3$, the U(1) 3-clock model becomes 
\begin{eqnarray}
\label{LE}
H_{\rm U(1)}&=&-\frac{J_x}{4}\sum_{i=1}^{L-1}
(S^+_i S^-_{i+1} + S^-_i S^+_{i+1}) \nonumber \\
&& +3h\sum_{i=1}^{L} (S^z_i)^2+{\sf const},
\end{eqnarray}
used by Luther and Scalapino \cite{Luther-Scalapino} as a quantum Hamiltonian 
approximation to the planar XY model. It has been established since then that this model shows 
a gapless phase and a phase transition to a gapped paramagnetic state for some 
value of \(h> 0\). However, this transition is {\it not} a BKT transition. Rather it is a 
Gaussian transition \cite{KM} similar to the dimerization transition of the AF
SU(2)-symmetric $J_1$-$J_2$ spin-$1/2$ chain, see Table\,\ref{MainTable}.  

As mentioned above, the conjugate representation depends on the parity of $p$.
There is also a difference in terms of symmetries of the U(1) clock model for \(p\) even
versus \(p\) odd.  Consider the following 
Hermitian matrix
\begin{eqnarray}
X=
\begin{pmatrix}
0& 0& 0& \cdots&  0 & 1\\
0& 0& 0& \cdots & 1 & 0\\
0& 0& 0& \cdots & 0& 0\\
\vdots& \vdots& \vdots&    & \vdots & \vdots \\
0& 1& 0& \cdots & 0 & 0\\
1& 0& 0& \cdots&  0 & 0\\
\end{pmatrix} , 
\end{eqnarray}
with the property $X^2=\one$. 
Its action on the elementary degrees of freedom depends only on the parity of $p$,
\begin{eqnarray}
X \tilde{U} X&=&\tilde{U}^\dagger \ \ , \ \  \ \ X \hat{V} X=\hat{V}^\dagger , \ \mbox{for $p \in {\sf odd}$} , \nonumber \\
X \tilde{U} X&=&\omega \tilde{U}^\dagger \ \ , \ \  X \hat{V} X=\hat{V}^\dagger , \ \mbox{for $p \in {\sf even}$}.
\end{eqnarray}
Then, for $p \in {\sf odd}$, $[H_{\rm U(1)}, \hat{X}]=0$, with $\hat{X}=\prod_{i=1}^L X_i$. 
Since $\hat{X}$ is hermitian and squares to the identity, it is a \(\mathbb{Z}_2\) symmetry 
of the \(U(1)\) clock models with odd \(p\). 
This symmetry, or the lack of it, may have implications for the quantum phase diagram of the U(1) clock model. 
The cases of  $p  \in {\sf odd}$ and $p  \in {\sf even}$, lead to different quantum phases and  types of transitions
(see Fig. \ref{fig:u1sketch}).

Finally,  notice that in the U(1) clock model the sign of the coupling $J_x$ is irrelevant for any $p$. 
This is due to the fact that if we act on the Hamiltonian (\ref{HU1}) with the unitary operator $W=\prod_{i \in \sf odd} 
\exp[-\im \pi S^z_i]$,
\begin{equation}
W^{\dagger}{H_{\rm U(1)}}W ,
\end{equation}
then $J_x \to - J_x$. This follows from 
Eq. (\ref{symcjh}) and the fact that $W$ commutes with the $\tilde U_i, \tilde U^\dagger_i$ operators.
For the $\mathbb{Z}_p$ clock model, however, the same transformation changes  $J_x \to - J_x$ only 
for even $p$. In this latter model, and for even $p$ (though not for odd $p$), one can further show 
that in addition the sign of $h$ can also be changed by the same unitary transformation.  

\subsection{Quantum Phase diagrams}

\begin{figure}[ht]
\includegraphics[width=8cm]{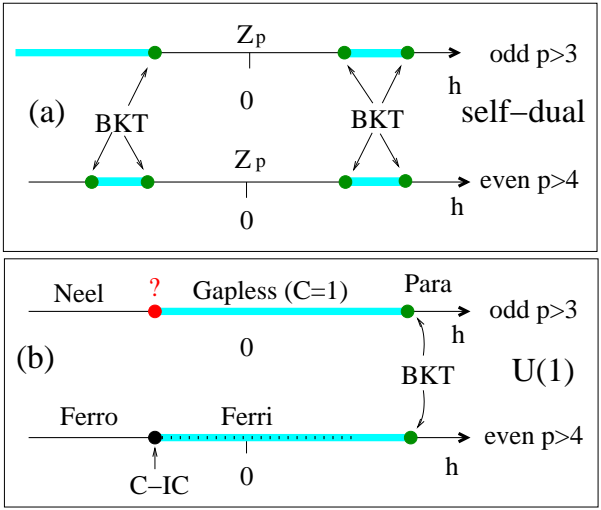}
\caption{Quantum phase diagram for (a) $\mathbb{Z}_p$ self-dual and (b) U(1)-symmetric 
$p$-clock chains. Gapless regions are denoted by a thick cyan line. With increasing $p\to \infty$ 
an  spontaneously $\mathbb{Z}_p$ symmetry broken region (with $p$-fold degenerate ground states) shrinks 
towards $h=0$. With increasing $p$ a Ferri phase for even $p$, as well as N\'eel to gapless 
transition points for odd $p$, shift towards $h=0$ in (b). }
\label{fig:u1sketch}
\end{figure}

Figure \ref{fig:u1sketch} (b) shows another main result of this paper: the quantum phase diagram of
our U(1)-symmetric $p$-clock chain, Eq. (\ref{HU1}), for odd and even values of $p \ge 5$. 
To emphasize the effect of the {\it explicit} U(1) symmetry and the parity 
of \(p\), we also present the quantum phase diagrams for the usual, self-dual clock model in Fig. \ref{fig:u1sketch} (a). 

Neither the \(\mathbb{Z}_p\) self-dual nor the U(1)-symmetric clock models show
BKT transitions for \(p<5\). Let us summarize the situation for the U(1) clock models.
The case \(p=2\) is simply the spin \(1/2\) isotropic XY model in a transverse field, 
see Table \ref{MainTable}. For $p=3$, the Luther-Scalapino model, there is no gapped 
phase for $h\to -\infty$, but rather a dominant 
nematic phase emerges and there is an Ising transition from the so-called XY1 to XY2,
the latter realized for large and negative values of $h$, gapless phases. In addition, 
the transition from the XY1 phase to the gapped paramagnetic phase is Gaussian rather than BKT. 
For $p=4$ the U(1) clock model seems to be new in the literature. We find direct C-IC 
transitions out of a gapless phase that we call a a Ferri phase into the gapped phases 
that appear for large \(|h|\) \cite{comment}.

\begin{figure}[ht]
\includegraphics[width=8cm]{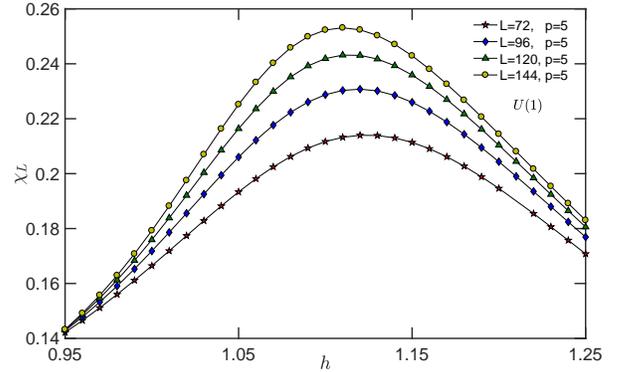}
\caption{Behavior of FS for the U(1) $p=5$ clock model for various system sizes.}
\label{fig:u1peakposition}
\end{figure}

Now for \(p\geq 5\) let us first look at large and positive values of $h$. As explained in detail in the next section,
the self-dual clock model displays an effective (emergent) U(1) symmetry in this regime. 
Hence, it is not surprising that the quantum phase 
diagrams look similar for large \(h\) for both the self-dual and U(1) clock models. In fact, preliminary
numerical values for the large-\(h\) BKT transition points in the U(1) clock model are very close
to those for the self-dual clock model for $p=5, 6$. In Fig. \ref{fig:u1peakposition} we show the behavior 
of the FS for the U(1) $p=5$ clock model for different system sizes. We observe a single peak 
near $h \approx 1$ and the peak position moves from the  massive side (large $h$ regime) towards 
the BKT point. Unfortunately we cannot extrapolate reliably the FS 
peak position to its thermodynamic value due  to problems similar to those already discussed for the \(\mathbb{Z}_p\) case with  $p>5$.

\begin{figure}[ht]
\includegraphics[width=8cm]{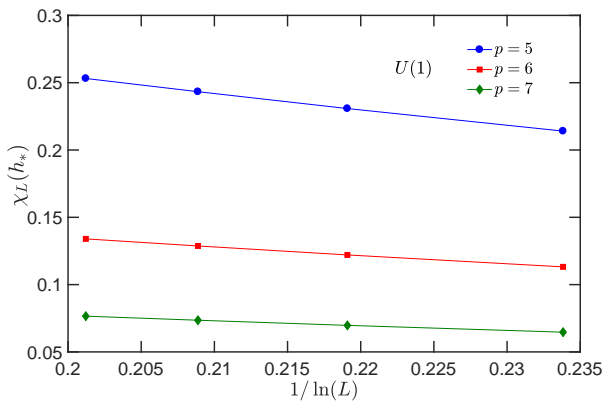}
\caption{Scalings of the FS peak heights with system size for U(1) $p$-clock chains at the critical boundary 
point between the gapless regime and paramagnetic disordered phase for $h>0$, confirm the BKT 
nature of the QPTs for $p=5, 6$, and $7$.}
\label{fig:u1fs}
\end{figure}
In Fig.  \ref{fig:u1fs} we present the scaling of the FS peak height with system size for 
the U(1) clock, confirming the BKT nature of its high-\(h\) phase transition for $p=5, 6$ and $7$.
Since the U(1) clock model is not self-dual, the nature of its other transitions must be settled
independently.

Let us consider next lower moderate values of \(h\). 
For the U(1) clock model and odd $p$, all ground states (for 
any value of $h/J_x$) have $\langle S^z_{T}\rangle=0$, while for even $p$ only the rightmost 
region of the gapless  phase and paramagnetic disordered phases have $S^z_{T}=0$. 
For $p  \in {\sf even}$ and larger than four, there appears a special regime inside the gapless 
phase that we call the Ferri phase. In this regime, the ground state expectation value of 
$S^z_{T}$ changes continuously with decreasing $h$ from  $\langle S^z_{T}\rangle=0$ 
to $|\langle S^z_{T}\rangle|=(p-1)/2$. For $p=4$ there is a direct C-IC transition from the 
gapless Ferri phase to  a gapped paramagnetic phase with increasing $h$ and no BKT transition.

A peculiar point is the transition from the gapped N\'eel state to the gapless phase for odd values 
of $p>3$, in which case different instabilities compete. For $p\ge 5$ and in the limit $h\to -\infty$ 
we can show by using an effective spin-1/2 model that  the ground state is a doubly-degenerate 
N\'eel ordered gapped state (see Appendix \ref{App A}). This is a simple consequence of the fact that Ising-like exchanges 
between the effective spin-1/2 variables emerge to second order in $J_x$, whereas exchange 
interactions emerge only in quartic or higher orders for $p>3$. The two ground states in the N\'eel 
phase are $| \frac{p-1}{2},\frac{1-p}{2}, \frac{p-1}{2},\frac{1-p}{2}, \dots  \rangle$  and 
$| \frac{1-p}{2},\frac{p-1}{2}, \frac{1-p}{2},\frac{p-1}{2}, \dots  \rangle$. Emergence of this 
doubly degenerate ground state for odd values of $p$ is a direct consequence of the 
$\mathbb{Z}_2$ symmetry discussed above, an exclusive property of odd-$p$ U(1) models.

At the boundary separating the N\'eel from the gapless phase no ground state level crossing is involved. 
However, there is a level crossing in the first excited state, and namely 7 different energy eigenstates 
cross at this point in thermodynamic limit. Among those 7 states:
\begin{itemize}
\item{ Two states have $S^z_{T}=\pm 1$. These are the lowest-energy excited states in the gapless region. }

\item{ Additional 4 lowest-energy excited states in the N\'eel gapped 
phase: Two states have $S^z_{T}=\pm (p-1)$, and the other two have $S^z_{T}=0$ (these 
4 states are made of pair of deconfined domain walls, where a single domain wall represents 
a $\pi$-soliton on top of the N\'eel configuration). }

\item{ The seventh state has $S^z_{T}=0$ and  becomes degenerate with the ground state 
in the N\'eel phase.}
\end{itemize}
Our numerical results support the scenario of a continuous second-order phase transition between the N\'eel and 
gapless phase with a central charge estimated to be $c \simeq 3/2$ (see Appendix \ref{App B}). 
However, we cannot make any definitive 
claims due to the poor convergence of our numerical results. The FS grows quite fast, namely as $L^2$,
for small systems $L\le 24$. To address the behavior of the FS for larger system sizes we are hindered by 
the very quick occurrence of a double degeneracy of the ground state in the N\'eel phase. Further studies are 
needed to determine the nature of this phase transition.

\section{Large $p$-limit: exact mapping to \\ a quantum rotor chain}
\label{largep}

In this section we study the large $p$-limit of the quantum self-dual $p$-clock chain. 
The partition function \({\cal Z}_p\) of the classical planar \(p\)-clock model approaches the partition function 
\({\cal Z}_{\rm XY}\) of the XY model in an obvious fashion as \(p\) grows. The situation is more subtle for
their corresponding $d=1$ quantum Hamiltonians. Classical partition functions describe the low-energy
physics of their associated quantum Hamiltonians. Hence, based on the relatively obvious relationship
\(\lim_{p\rightarrow \infty}{\cal Z}_p={\cal Z}_{\rm XY}\), we can at best expect that the quantum $p$-clock 
chain approaches the quantum planar-rotor chain as \(p\) grows to infinity only in the low-energy sector. This is true,
in general, of any two quantum Hamiltonians with associated partition functions that 
match in some limit. There is, however, a twist to this expectation in the present case. 

\begin{figure}[thb]
\includegraphics[width=7cm]{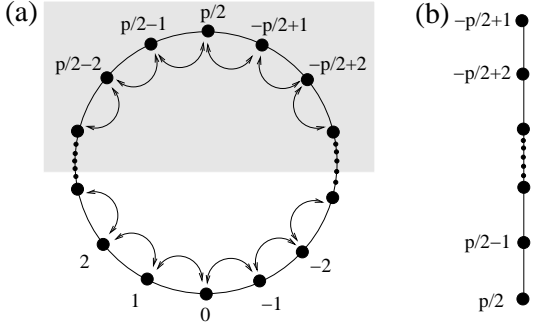}
\caption{(a) Allowed $n_j$ values for even $p$. Arcs with arrows at the ends connect $n_j$'s  
related by a single action of $H_1$. We have shaded quantum numbers $|n_j|= {\cal O}(p)$, which can be safely 
truncated when considering ground state (low energy) physics for $p\to \infty$ as proved in the text. (b) When 
quantum numbers $n_j$ are defined on a segment, instead of a circle, the system displays an explicit $O(2)$ symmetry 
for any $p>2$ (for $p=3$ the model becomes the spin-1 chain given in Eq. (\ref{LE})).}
\label{fig:AllowedQuantumNumbers} 
\end{figure}

Let's look at the Hamiltonians in question more closely. The effective Hamiltonian associated to
\({\cal Z}_{\rm XY}\) is the quantum planar-rotor chain 
\begin{eqnarray}
\label{XYham}
H_{\sf XY}= \sum_{j}\left [\frac{1}{2} \hat L_j^2-\lambda \cos(\hat \theta_{j+1}-\hat \theta_j)\right ],
\end{eqnarray}
with \(\hat L_j=-\im\partial_{\theta_j}\) and 
\(2\cos(\hat \theta_{j+1}-\hat \theta_j)=U_i U_{i+1}^\dagger+U_i^\dagger U_{i+1}\). 
We can re-write the familiar quantum \(p\)-clock chain in a way that facilitates comparison
by introducing a Hermitian momentum-like operator $\hat{p}_{\theta}$ as
 $V=\exp[\im \epsilon_p \hat{p}_{\theta}]$ with \(\epsilon_p=2\pi/p\).
This relationship does not specify $\hat{p}_{\theta}$ completely, but we can complete
the specification by demanding that 
 the quantum numbers associated to 
 $\hat p_{\theta_j}$ are given 
 by the integers 
\begin{eqnarray}
\label{forsos}
n_j = \left \{ 
\begin{array}{cc}
-\frac{p}{2}+1,\ldots, \frac{p}{2} &\,\,\mbox{for $p$ even} \\
& \\
- \frac{p-1}{2},\ldots, \frac{p-1}{2} & \,\, \mbox{for $p$ odd} \\
\end{array} \right . .
\end{eqnarray}
See Fig. \ref{fig:AllowedQuantumNumbers} for even $p$.
Similarly, let us introduce  $\hat \theta$ such that $ U=\exp [ \im  \hat \theta]$. Then,
\begin{equation}
\label{modelD}
H_{\sf VP}=-  J_x\sum_{j}\cos{( \hat \theta_j -\hat \theta_{j+1})} -2h\sum_{j}\cos(\epsilon_p\hat{p}_{\theta_j}) .
\end{equation}

Comparing Eqs.\,\eqref{XYham} and \eqref{modelD} one wonders whether the
low-energy properties of these Hamiltonians may match as \(p\rightarrow \infty\). On
one hand, the nearest-neighbor interaction term  for the \(p\)-clock does approach
the corresponding term of the rotor chain uniformly. On the other hand, the spectra of the operators
\( \cos(\epsilon_p\hat{p}_{\theta_j})\) and \(\hat L_j^2\) remain completely different even for arbitrarily
large \(p\). This is precisely the reason why the self-duality of the \(p\)-clock chain is not shared
by the planar-rotor chain. Hence, loosely speaking, if the control parameters are such that the ground state of the
clock model fluctuates over all values of \( \cos(\epsilon_p\hat{p}_{\theta_j})\), one cannot expect
that the ground state properties of the \(p\)-clock and the rotor chain agree no matter how large \(p\)
grows. The ground state properties of the two Hamiltonians can only match in the large-\(p\) limit
if we impose some constraints on the control parameters of the \(p\)-clock model.  

The strength of quantum fluctuations (``transverse field'' $h$) is related to the temperature 
of the corresponding classical planar $p$-clock models in the limit $J_y/T \to \infty$
\begin{equation}
\label{transl}
h=Ta_1,
\end{equation} 
where $a_1=a_1(T/J_y,p)$ satisfies the exact relation obtained by utilizing bond-algebraic dualities \cite{Ortiz},

\begin{equation}
{ \partial_{a_1} \ln \left[ \sum_{s=0}^{p-1}e^{2a_1\cos(\epsilon_p s)}\right]}=
2 \, e^{\frac{J_y}{T}\left[\cos (\epsilon_p) -1\right]}.
\label{dual}
\end{equation}
We have put above $k_B=1$, hence temperature is measured in units of energy $J$.

In the large-$p$ limit, Eq. \eqref{dual} reduces to \cite{Ortiz} $a_1(T/J)= T/(2J_y \varepsilon^2_p)$, and the 
quantum $p$-clock chain Hamiltonian in variables $\hat \theta$ and $\hat p_{\theta}$ becomes
($2 \eta =-\left (\varepsilon_p J/T\right )^2 <0$)
\begin{equation}
H_{\sf VP}=\!-J_x\!\sum_{j} \left [
\frac{1-\cos (\varepsilon_p \hat p_{\theta_j})}{2\eta}+\cos (\hat \theta_j-\hat \theta_{j+1}) \right ] , 
\label{modelplarge}
\end{equation}
where $\cos (\hat \theta_j-\hat \theta_{j+1})=(U_{j+1}^\dagger U^{\;}_j+ U_{j}^\dagger U^{\;}_{j+1})/2$.

We next show the mathematical conditions under which one can replace the 
$1-\cos(\varepsilon_p \hat p_{\theta_j})$ term in Eq. \eqref{modelplarge} by $\varepsilon^2_p \hat{p}^2_{\theta_j}/2$, 
in the large-$p$ limit. We start by establishing a simple upper bound on the ground state energy $E_0$ of 
$H_{\sf VP}$ with the help of the product states $\ket{{\bf n}}=\ket{n_1,n_2,\cdots, n_L}$. 
In particular, choosing $\ket{\Psi_v}=\ket{0,0,\cdots,0}$ it results  $E_0 \leq \bra{\Psi_v} \hat H_p \ket{\Psi_v}=0$. 
Furthermore, since 
$\langle \Psi_0| \cos(\hat \theta_j-  \hat \theta_{j+1})|\Psi_0\rangle \le 1$, with $|\Psi_0 \rangle$ the 
ground state of $H_{\sf VP}$, one obtains the following inequality
\begin{equation}
\label{inequality}
0\le \langle \Psi_0| 1-\cos (\varepsilon_p \hat p_{\theta_j}) |\Psi_0\rangle \le -2\eta .
\end{equation}

If one can prove that, for $\eta \to 0$, the quantum numbers $|n_j|={\cal O}(p)$ get projected out of 
the low-energy sector of  Hamiltonian \eqref{modelplarge} (as depicted in Fig. \ref{fig:AllowedQuantumNumbers}(a)), 
then the harmonic approximation to $1-\cos (\varepsilon_p \hat p_{\theta_j})\to (\varepsilon_p \hat p_{\theta_j})^2/2$ becomes 
exact  for that sector. To prove this, we re-scale Hamiltonian \eqref{modelplarge} as
\begin{equation}
\label{eqtilde}
\tilde H= -2\eta \hat H_{\sf VP}/J_x=  H_0+\eta H_1,
\end{equation}
with $H_1=2\sum_{j} \cos(\hat \theta_j -\hat \theta_{j+1})$, and ground state energy $\tilde E_0=-2\eta E_0/J_x$. Since 
$\langle \Psi |\eta H_1|\Psi \rangle$ ($\langle \Psi |\Psi \rangle=1$)
cannot be less than $2\eta L$ and $\langle \Psi |H_0|\Psi \rangle$ cannot become negative, 
the ground state energy density of $\tilde H$ must satisfy 
\begin{equation}
2\eta \le \tilde E_0/L \le 0 ,
\end{equation} 
where the upper bound results from considering the ground state of $H_0$, $\ket{\Psi_v}$.

Consider the expansion of the ground state of $\tilde{H}$
\begin{eqnarray}
\label{exp}
\ket{\Psi_0}&=&\sum_{{\bf n}} \alpha_{{\bf n}} \, \ket{{\bf n}}, \ \  \sum_{{\bf n}}|\alpha_{{\bf n}}|^2=1, 
\end{eqnarray}
in terms of eigenstates of $H_0$, i.e., $H_0 \ket{{\bf n}}=\epsilon_{{\bf n}}\ket{{\bf n}}$. In the following we will denote by 
$\ket{{\bf m}}$ the basis states in the expansion of Eq. (\ref{exp})  where  at least one quantum number $|n_{j}|={\cal O}(p)$.
 Our goal is to show that in $p\to \infty$ limit all such 
basis states $\ket{{\bf m}}$ do not contribute to the expansion of $\ket{\Psi_0}$, i.e. $\sum_{{\bf m}}| \alpha_{{\bf m}}|^{2} \to 0$.

Let us focus on one such concrete state  $\ket{{\bf m}_0}=\ket{m_1,m_2,\cdots, m_L}$, with one quantum number 
$|m_{j=\ell}|=\tilde n={\cal O}(p)$ (if $\ket{{\bf m}_0}$ has several quantum numbers with $|m_{j}|={\cal O}(p)$ we 
will assume that $|m_\ell|\ge |m_j|$ for all other $j$'s). 
Then, $\epsilon_{{\bf m}_0} \ge 1-\cos{(\varepsilon_p \tilde n)} \gg \frac{1}{p}$.

Starting from $\tilde H \ket{\Psi_0}=\tilde E_0\ket{\Psi_0}$ one can write the amplitude 
$\alpha_{{\bf m}_0}=\bra{{\bf m}_0}\Psi_0\rangle$, knowing that  $2\eta L \le \tilde E_0\le 0$ and hence 
$ \epsilon_{{\bf m}_0}\gg |\tilde E_0|$, as
\begin{equation}
\label{amplitudes}
\alpha_{{\bf m}_0} = - \frac   {\eta+ {\cal O}(L \eta ^{2})}{\epsilon_{{\bf m}_0} } 
\sum_{j=1}^L(\alpha_{{\bf m}_{0j}^-}+\alpha_{{\bf m}_{0j}^+}) ,
\end{equation}
where $\alpha_{{\bf m}_{0j}^\pm} \! = \! \langle m_{1},\cdots \!,m_{j}\! \pm \! 1, m_{j+1}\! \mp \!1,\cdots \!, m_{L}|\Psi_0\rangle$ 
are amplitudes  of those eigenstates of $H_0$ which contribute to the ground state expansion in Eq. (\ref{exp}) and acting 
on which by $H_1$ gives $\ket{{\bf m}_0}$ (there are at most $2L$ such amplitudes for periodic boundary conditions) and 
in their own they satisfy analogous equations.

From Eq. (\ref{amplitudes}) and since $|\eta+{\cal O}(L \eta ^{2})|<2|\eta|$ and $\epsilon_{{\bf m}_0} \gg 1/p$, 
we obtain the following estimate,
\begin{eqnarray}
\label{inequality001}
|\alpha_{{\bf m}_0}| &<&\! 4Lp|\eta \alpha_{{\bf m}_1} | ,
\end{eqnarray}
where we have denoted by $\alpha_{{\bf m}_1}$ the amplitude from the set of $\{\alpha_{{\bf m}_{0j}^\pm} \}$ with 
maximal modulus, $|\alpha_{{\bf m}_1}|= \max \{| \alpha_{{\bf m}_{0j}^\pm}  |\} $.

Similarly for $|\alpha_{{\bf m}_1}|$ we can derive analogous to inequality (\ref{inequality001}) estimate, 
$|\alpha_{{\bf m}_1}| < 4Lp|\eta \alpha_{{\bf m}_2} |$, where  $|\alpha_{{\bf m}_2}|= \max \{| \alpha_{{\bf m}_{1j}^\pm}  |\}$.
As long as state contains at least one quantum number with $|n_j|={\cal O}(p)$ (and hence for 
$p\to \infty$ its corresponding  $\epsilon_{{\bf m}_1}\gg 1/p$) we get the following rigorous bounds, 
\begin{eqnarray}
\label{inequality1}
|\alpha_{{\bf m}_0}| &<&\! (4L |\eta| p)  |\alpha_{{\bf m}_{1}} |< \!(4L |\eta| p)^{2}
|\alpha_{{\bf m}_{2}}|\nonumber\\
&<&\cdots<(4L |\eta| p)^{\frac{\tilde{n}}{2}}  |\alpha_{{\bf m}_{\tilde{n}/2}}|.
\end{eqnarray}
In deriving the above chain of inequalities we observed that all the states obtained from 
$|{\bf m}_0\rangle$ by acting with $H_1$ once, twice, $\cdots, \tilde{n}/2$ times ($H_1|{\bf m}_0\rangle$, 
$H^{2}_1|{\bf m}_0\rangle,\cdots,H^{\tilde{n}/2}_1|{\bf m}_0\rangle $) contain at least one quantum number 
(at site $j=\ell$) with absolute value $|n_\ell|\ge \tilde{n}/2={\cal O}(p)$.
Since any amplitude in the expansion of Eq. (\ref{exp}) satisfies $|\alpha_{\bf n}| \le 1$, the inequality (\ref{inequality1}) 
implies $|\alpha_{{\bf m}_0}| <(4L |\eta| p)^{\frac{\tilde{n}}{2}}$ and thus,
\begin{equation} 
\label{sqa}
\sum_{\bf m} |\alpha_{\bf m}  |^2 < \sum_{\bf m} (4 L |\eta| p)^{\max\{|n_{j}|\}}.
\end{equation}
Since for $p\to \infty$, $4 L |\eta| p\to 0$, we can change the power $\max  \{ |n_{j}|\} $ (which from the 
definition of states  $\ket{{\bf m}}$ are of order ${\cal O}(p)$) to an order of magnitude smaller number, e.g. 
$\sqrt{p}$, without violating the inequality, in fact, strengthening.
We can then bound the number of summands  by the total 
number of states, $p^{L}$, and obtain a very modest but exact upper bound for the left-hand-side of inequality (\ref{sqa}), 
\begin{equation} 
\label{sqa1}
 \sum_{\bf m} |\alpha_{\bf m}  |^2   < p^{L-\sqrt{p}} \Big(\frac{8\pi^{2}J^{2}L}{T^{2}}\Big)^{\sqrt{p}}.
\end{equation}
Thus, as we agreed to take $p\to \infty$ limit first before taking the thermodynamic limit, inequality 
(\ref{sqa1}) proves that the subspace of eigenstates of $H_0$ which contains at least one quantum number 
with $|n_j|={\cal O}(p)$ (see shaded region in Fig. 2 (a)), becomes projected out of the ground state 
\cite{comment0} of $\tilde H$ and thus of $\hat H_p$.

Taking $p\to \infty$ limit at finite $T/J>0$ in the right-hand-side of Eq. \eqref{modelplarge}, using 
$z^2[1-\cos{(A/z)}]=A^2/2[1+{\cal O}(A^2/z^2)]$ for $A/z\to 0$, as far as low energy physics is concerned, we 
obtain the following exact representation of the quantum $p$-clock chain Hamiltonian,
\begin{equation}
\label{fin}
\hat H_p=J_x\sum_{j}    \left[  \frac{g}{2} \hat p^2_{\theta_j} -\cos(\hat \theta_j-  \hat \theta_{j+1}) \right],
\end{equation}
where the coupling constant $g$, the strength of quantum fluctuations, is related with the temperature of the 
classical 2d extremely anisotropic $p\to \infty$-clock model as 
\begin{equation}
\label{main}
g=\frac{T^2}{J_xJ_y}.
\end{equation}

At this point let us recall the matrix representation of $\hat p_{\theta}$ and $U$ in the 
eigenbasis of $\hat p_{\theta}$ (e.g., for even $p$), 
where $\hat p_{\theta}=\mathrm{diag}(-\frac{p}{2}+1,\cdots,\frac{p}{2}-1,\frac{p}{2})$, and 
\begin{eqnarray}
{U}=
\begin{pmatrix}
0& 0& 0& \cdots& 1\\
1&0 & 0& \cdots& 0\\
\vdots& \vdots& \vdots&      & \vdots \\
0&  \cdots&1& 0&0\\
0& 0& \cdots& 1&0\\
\end{pmatrix}.
\end{eqnarray}

If we calculate the commutator between $\hat p_{\theta_j}$ and $ U_j$ restricted to the subspace where quantum numbers 
$|n_j|={\cal O}(p)$ are discarded as indicated in Fig. \ref{fig:AllowedQuantumNumbers}(a) (in fact we must restrict to this 
subspace since our Hamiltonian \eqref{fin} was obtained for this very subspace) we obtain,
\begin{equation}
\label{commutator}
[\hat p_{\theta_j},  U_l]   = \delta_{jl} U_j,
\end{equation}
which implies that $(\hat \theta, \hat p_{\theta})$ act as canonically conjugate variables, i.e., 
$[\hat \theta_l, \hat p_{\theta_j}]= \im\delta_{jl}$. 
Therefore, in the $p\rightarrow \infty$ limit, and within the low-energy sector, 
$\hat p_{\theta_j} \rightarrow -\im \partial_{\theta_j}$ and the 
Hamiltonian (\ref{fin}) is recognizable as the one describing the quantum rotor chain. 

A reasoning similar to the one used to obtain inequality (\ref{inequality}) leads to 
$\Delta \hat p_{\sf rotor}^2(g)= \!\!\langle \Psi_0(g)| -\frac{\partial^2}{\partial \theta^2_j}|\Psi_0(g) \rangle \le \frac{2}{g}$, where 
$|\Psi_0(g) \rangle$ denotes the ground state of the quantum rotor chain for coupling strength $g$. Thus, 
$\Delta \hat p_{\sf rotor}^2$ is small even at the BKT transition point: $ \Delta \hat p_{\sf rotor}^2 (g_c) \lesssim 2.5$. 
We note that the emergent $O(2)$ symmetry in the low-energy sector of the quantum $p$-clock chains finds a natural 
interpretation for the regimes where the momentum quantum numbers $|n_j|={\cal O}(p)$ decouple from the 
low-energy physics. This is so because one cannot distinguish between the case where the $n_j$-s are defined 
mod $p$ ($\mathbb{Z}_p$ discrete symmetry as in Fig. \ref{fig:AllowedQuantumNumbers}(a)) 
from the case where $n_j$-s are defined on a segment, the latter case displaying an explicit $O(2)$ symmetry for any $p>2$.   

\section{Estimating the BKT temperatures of classical planar clock-models}
\label{classicalmap}

In this section we compare our results on the QPTs of the self-dual $p$-clock chains with the 
finite-temperature  phase transitions of the classical two-dimensional $p$-clock model.
The Hamiltonian of the classical \(p\)-clock model is 
\begin{eqnarray}
H_{2d}= -\sum_{\langle {\bs r}, {\bs r'}\rangle } J_{{\bs r}, {\bs r'}} \cos(\theta_{\bs r}-\theta_{\bs r'}).
\label{XYmodel}
\end{eqnarray}
The angles $\theta_{\bs r}=2\pi s_{\bs r}/p$ live on the square lattice and take on 
discrete values, determined by the set $s_{\bs r}=0,1,\ldots, p-1$ (see Fig. \ref{fig:clock}). 
The classical 
$p$-clock model interpolates between the two-dimensional classical Ising model 
($p=2$) with $\mathbb{Z}_2$ symmetry and the  two-dimensional XY model ($p=\infty$) with 
continuous $O(2)$ symmetry.
In the following we will denote the exchange couplings  along horizontal bonds by $J_{{\bs r},{ \bs r +\vec{e}_x} }=J_x$ 
and along vertical bonds by $J_{{\bs r},{ \bs r +\vec{e}_y} }=J_{y}$ and let $\sqrt{J_{x} J_{y}}=J$.
It was predicted analytically that the two-dimensional classical $p$-clock model, for large enough $p$,  
possesses a couple of temperature induced topological BKT phase transitions separating 3 distinct 
phases \cite{Kadanoff, Elitzur}: a low temperature ordered phase with spontaneously broken discrete 
$\mathbb{Z}_p$ symmetry for $T<T_{\sf BKT}^{(1)}$, a critical intermediate phase for 
$T_{\sf BKT}^{(1)}\le T\le T_{\sf BKT}^{(2)}$, and a disordered, high-temperature phase for 
$T>T_{\sf BKT}^{(2)}$.

Numerical studies of criticality in $p$-clock models have been mainly performed for
the classical model in two dimensions and in the isotropic regime. The most popular methods
are classical Monte-Carlo \cite{Lapilli,Baek, Borisenko, Oshikawa, Okabe,Surugan}, 
density-matrix \cite{Chatelain}, corner-transfer-matrix and tensor renormalization group 
\cite{Krcmarr, GC} simulations. Early Monte-Carlo studies \cite{Lapilli,Baek,Borisenko} raised certain 
controversies regarding these BKT transitions. In particular, the universality of the jump in 
the helicity modulus \cite{Fisher,Min} at these transitions, that is, its independence of $p$,
was brought into question. In later work,
the controversy was suggested to be resolved by computing numerically the helicity modulus 
defined with respect to a finite, quantized twist matching the discrete $\mathbb{Z}_{p}$ symmetry 
of the  model \cite{Oshikawa,Surugan}. 

The ground state behavior of one-dimensional quantum systems can be mapped, typically
in an approximate way, 
to the finite-temperature equilibrium behavior of two-dimensional classical statistical 
systems. This quantum-to-classical correspondence was pioneered by Feynman in the form
of path integrals. Quantum $p$-clock chains can be mapped in an exact way to 
two-dimensional classical $p$-clock models in a certain limit\cite{Ortiz}. The two 
systems should manifest similar universal (critical) behavior and there exists an exact 
mapping between the strength of the quantum fluctuations of the self-dual quantum $p$-clock 
chain and the temperature of the corresponding classical planar model. Because of 
this mapping we can determine numerically the critical BKT temperature,
$T_{\sf BKT}$, of the extremely anisotropic planar $p$-clock model.

With the help of the self-dual relations (\ref{transl}) and (\ref{dual}), \cite{Ortiz} one 
can translate the BKT transition points $\lambda^*_{\pm}$ of the quantum \(p\)-clock chain,
see Sec.\,\ref{FSpgeq5}, into BKT transition temperatures of the corresponding 
two-dimensional classical $p$-clock model. By the nature of the standard quantum-to-classical
mapping, our estimates should work best in the extreme anisotropic limit when $J_y/T\to \infty$, 
$J_x/T\to 0$ and $J_y/J_x \to \infty$. 
\begin{itemize}
\item
For $p=5$, we obtain  $T_{\sf BKT}^{(1)}\simeq 0.913J$ and 
$T_{\sf BKT}^{(2)}\simeq 0.946J$. The classical Monte-Carlo estimates \cite{Oshikawa, Surugan} 
of the critical temperatures in the isotropic regime $J_x=J_y$ are
$T_{\sf I,BKT}^{(1)}\simeq 0.908J$ and $T_{\sf I,BKT}^{(2)}\simeq 0.944J$. 
\item
For $p=6$, we obtain $T_{\sf BKT}^{(1)}\simeq 0.697J$ and $T_{\sf BKT}^{(2)}\simeq 0.906J$.
The classical Monte-Carlo estimates are $T_{\sf I, BKT}^{(1)}\simeq 0.700(2)J$ and 
$T_{\sf I,BKT}^{(2)}\simeq 0.904(2)J$, again for the isotropic \(p=6\)-clock model.
\end{itemize}
We conclude that the 
critical BKT temperatures of the extremely anisotropic and isotropic $p$-clock models for $p=5$ and 
$p=6$ are very close to each other. 
We note that for $p\to \infty$ both $T_{\sf BKT}^{(1)}$ and $T_{\sf I, BKT}^{(1)}$ approach zero as 
$O(1/p^2)$,\cite{Ortiz} and $T_{\sf I, BKT}^{(1)}$, in this limit, should coincide with the corresponding 
critical temperature of the self-dual Villain model \cite{Surugan}.

\section{Summary and Conclusions}
\label{Conclusion}

In two dimensions, the classical $O(3)$ Heisenberg model shows no phase transition at finite temperature, 
and discrete symmetry models of magnetism realize a variety of universality classes well
described within the Landau-Wilson theory of critical phenomena \cite{book}. The $O(2)$ 
XY model, is the oddball 
in between: there is a phase transition, the celebrated BKT transition, but because of 
Mermin-Wagner theorem this transition must necessarily occur with no associated local order parameter.
 
The BKT transition is at the center of a seemingly perennial attention for two reasons. On one hand, 
in spite of the exotic features of the BKT transition in the XY model,
the BKT transition is apparently the most common critical behavior
in two dimensions closely followed by the Ising universality class\cite{Taroni2008}.
On the other hand, it remains challenging to predict the appearance and/or detect 
BKT transitions in model systems.

In this paper we exploited the fidelity susceptibility (FS), a type of indicator
of the second-order rate of change of the ground state with driving parameter, as 
the numerical tool for detecting and identifying BKT transitions. While the FS is also suitable for identifying
generic critical points, it was understood only recently that the finite-size scaling of the 
FS obeys very precise and distinctive behavior at a BKT transition. Hence, by monitoring the 
behavior of the FS, we have confirmed 
in large-scale DMRG numerical simulations that the quantum self-dual $p$-clock chains for $p=5,6,7$ 
supports two BKT phase transitions located symmetrically around the self-dual point. This
result is important because the power of the FS to detect and identify BKT transitions
was tested before only in models with an explicit U(1) symmetry. In addition, using an exact 
quantum-to-classical correspondence, enabled by the bond-algebraic approach to the duality realized in 
the quantum $p$-clock chains, we determined the critical temperatures of the BKT transitions
in the classical two-dimensional $p=5,6$-clock models on the square lattices in the limit of 
extreme spatial anisotropy.

As mentioned above, the quantum \(p\)-clock chains with \(p\geq 5\) have two-phase transitions 
and are self-dual. As a result, it is enough to determine that only one of the two transitions
is  BKT. The other transition will then necessarily be BKT by self-duality. One
natural way of approaching this problem, that is, to ascertain that one of the two transitions
in the quantum \(p\)-clock chain is BKT, is to see whether one can map the ground state properties of 
the quantum \(p\)-clock chain into those of the quantum $O(2)$ rotor chain. 
In this paper we described a mathematically well-controlled procedure for
taking the limit $p\rightarrow \infty$ and showed that we do indeed recover
the quantum $O(2)$ rotor chain from the \(p\)-clock, as far as ground-state properties go. The
key to the success of this calculation is to appreciate that the limit $p\rightarrow \infty$
cannot yield the $O(2)$ rotor model everywhere in parameter space, but only
if one adjusts the control parameter of the quantum \(p\)-clock chain appropriately as \(p\) grows. 

Finally, in an effort to understand better what are the sufficient conditions for the appearance
of a BKT transition, we introduced and investigated a new class of clock model,  the U(1)-symmetric
quantum \(p\)-clock chain. We derived this model from the usual self-dual clock model
by restoring a natural continuous U(1) symmetry that
is explicitly broken in the quantum $p$-clock chain.  
The gapless phase broadens dramatically for the U(1)-symmetric model, the location of one 
of the BKT transition points is practically unaffected by restoring the continuous symmetry, 
and the other BKT point is replaced by non-BKT transitions whose character depends on the 
parity of \(p\). This picture is consistent with the supposition that presence of 
a gapless region and BKT point in the quantum \(p\)-clock chain are related to an emergent 
continuous symmetry \cite{Ortiz}, while the other BKT transition is forced in by the self-duality of the
model.

In a sense made precise in this paper we analyze the interplay between angular momentum
and Weyl algebras, the \(p\)-clock model is a close relative of many Potts models with 
explicit U(1) symmetry and a critical phase. Due to self-duality and other
symmetries of the \(p\)-clock model, there is more than one natural way of restoring an
explicit U(1) symmetry and none of them preserves the self-duality symmetry. Our  
numerical investigations suggest that our U(1) clock model is essentially a very special and unique 
way to restore a U(1) symmetry in the \(p\)-clock model while preserving the BKT character of at least 
one transition point for all $p \ge 5$. The fact that we were able to identify this model quickly among all the 
possibilities is a tribute to the power of the FS to diagnose BKT phase transitions.

\section{Acknowledgments}

We thank E. Fradkin for useful correspondence.
Computations were performed on the computational cluster at Nanjing University of Aeronautics 
and Astronautics and the Big Red II HPC cluster at Indiana University Bloomington. 
Most of the code was written in MATLAB. G.S. is 
appreciative of support from the NSFC under the Grant No. 11704186 and the startup Fund 
of Nanjing University of Aeronautics and Astronautics under the Grant No. YAH17053.

\appendix

\section{Effective spin-1/2 model}
\label{App A}

Here we derive an effective spin-$1/2$ model for the U(1) $p=5$  clock model for 
negative values of $h$ in the limit $|J_x/h|\to 0$. A similar derivation holds for other odd-$p>5$ cases. 

Let us start rewriting Eq. \eqref{HU1} as
\begin{equation}
H_{\rm U(1)}=H_0+\lambda H_1,
\end{equation}
where $h=-1/2$, $\lambda=-J_x/2$, and 
\begin{equation}
H_0=\sum_{i=1}^L\cos \left (\frac{2\pi n_i}{5} \right ),
\end{equation}
with $n_i=-2,-1,0,1,2$.
When $\lambda=0$ the ground state  is $2^L$-fold degenerate as, at each site $i$, $n_i$ 
can be in one of the two states $\pm 2$ which we will denote by $\ket{\! \uparrow}$ and $\ket{\! \downarrow}$, respectively.
We next switch $H_1$ on, and apply usual perturbation theory assuming an infinitesimally small $\lambda$. 
There is no contribution to first order in $\lambda$. The first non-trivial contribution is
second order. To second order,  only a $\sigma_i^z \sigma_{i+1}^z$ type interaction 
is generated and no $\sigma_i^{+} \sigma_{i+1}^-$ exchanges are allowed. This is so, 
since $\hat V_i^4 \ket{\! \uparrow}=\ket{\! \downarrow}$. Thus, one has to act 4 times with 
$H_1$ to flip neighboring spins. 

In summary, Ising-like exchanges appear in second order of 
$\lambda$, whereas spin-flip exchanges appear only in fourth order of $\lambda$. Now, 
without detailed calculation it is immediate that the sign in front of $\sigma_i^z \sigma_{i+1}^z$ is 
antiferromagnetic. This is so because the antiferromagnetic state is not an eigenstate of the total Hamiltonian.
Hence, it can lower the ground state energy by quantum fluctuations (as opposed to a ferromagnetic state that is a classical state). 

\section{Estimating the central charge}
\label{App B}

\begin{figure}[ht]
\includegraphics[width=8cm]{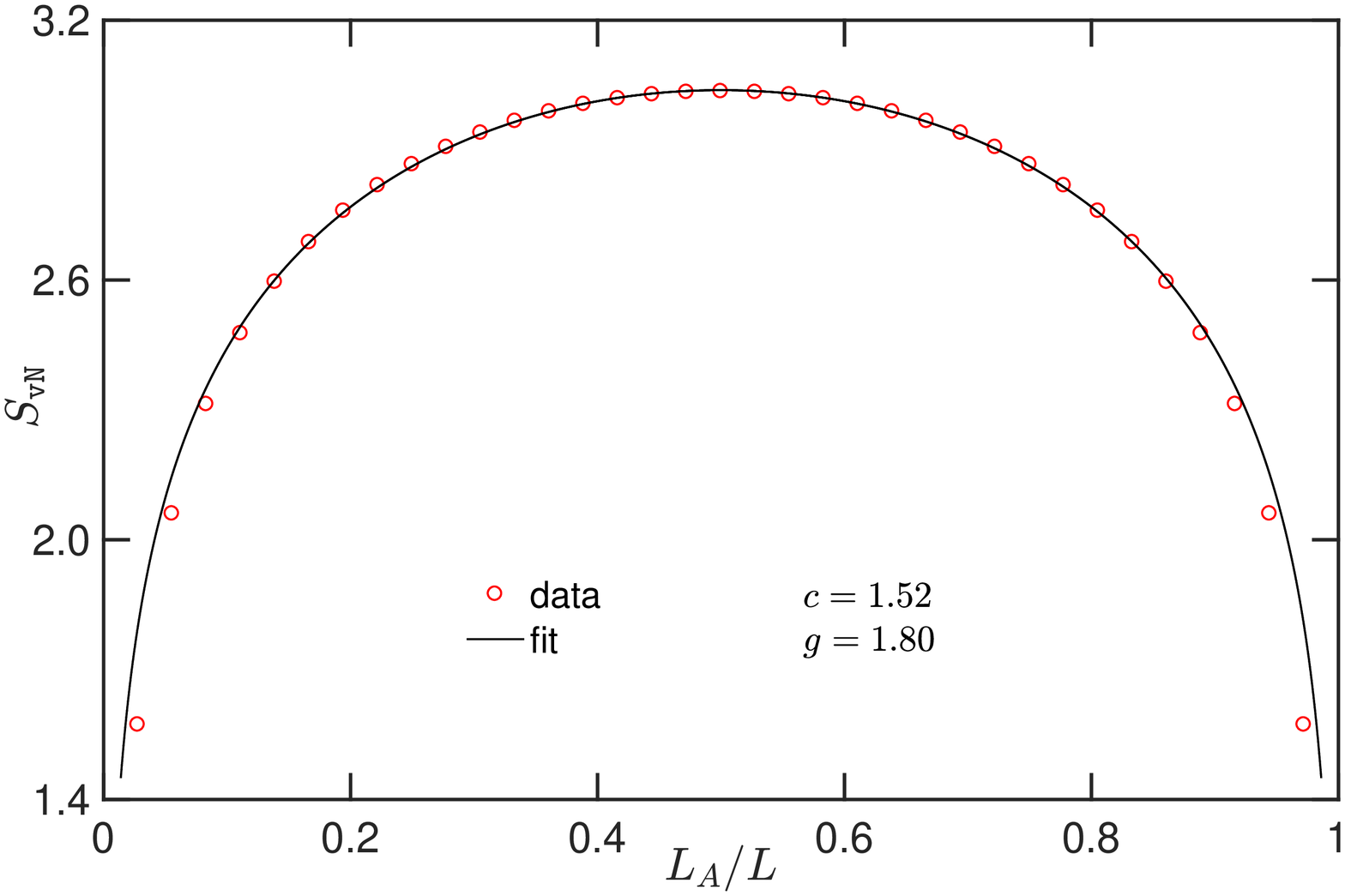}
\caption{Von Neumann entropy $S_{\rm vN}$ as a function of the partition size $L_A$ for system size $L=36$.}
\label{central_charge}
\end{figure}

Assume that  for negative values of $h$ the phase transition between the gapless and gapped (N\'eel) 
phases of the U(1) odd-$p$ clock model is of second order and described by some conformal field theory. 
Then we can estimate  \cite{Kitaev,Cardy} the central charge $c$ associated with this phase transition from the von 
 Neumann entanglement entropy of subsystem $A$
\begin{equation} \hspace*{-0.1cm}
\label{Kitaev}
S_{\rm vN}=-{\rm Tr} \big[ \rho_A\ln(\rho_A)   \big]=\frac{c}{3}\ln  \left [ \frac{L}{\pi } \sin \left ( \pi \frac{L_A}{L} \right)  \right] + g ,
\end{equation}
where $g$ is a non-universal constant  and $\rho_A$ is the reduced density matrix of  subsystem $A$ of length $L_A$ in a bipartition of the full system into two parts of linear sizes $L_A$ and $L-L_A$. 
Figure \ref{central_charge} displays a fit of the computed $S_{\rm vN}$ to the analytical result of Eq. (\ref{Kitaev}), by assuming $g\simeq 1.8$  
and $c\simeq 1.52 \simeq 3/2$. Note that we can only simulate small system sizes since by increasing the 
system size the entanglement entropy becomes so large that  it invalidates our accuracy. 



\begin{thebibliography}{10}
\bibitem{foot}
One can show by various arguments, for example, a Peierls-type argument\cite{Ortiz},
that the ferromagnetic phase must shrink 
as \(p\) increases, that is, \(\lim_{p\rightarrow\infty}\lambda_-^*=0\). In addition,
because of the self-duality symmetry of the \(p\)-clock chain, the width of the critical 
phase is \(\lambda_+^*-\lambda_-^*= \frac{1-\lambda_-^{*\, 2}}{\lambda_-^*}\).
\bibitem{comment}  
Numerically, for large and negative values of $h$, we observe a saturation of the ground state's 
quantum number associated with the total $S^z$ to a maximum allowed value. The ground state 
value of the total $S^z$ changes continuously in the gapless phase (so-called Ferri phase) and it saturates 
to a maximum value in the gapped phase. There is no jump of $S^z$ involved, the transition is a 
standard C-IC phase transition.
\bibitem{comment0} 
We note that exactly the same reasoning as used for the ground state of $\tilde H$ can be used to prove that the subspace 
of eigenstates of $H_0$ which contains quantum numbers $|n_{j}|={\cal O}(p)$ becomes projected out of all low-energy 
states of $\tilde H$ (and hence of $\hat H_p$) as well for $p\to \infty$.
\end{thebibliography}
\end{document}